\documentclass[aps,pra,twocolumn,letterpaper,superscriptaddress,10pt]{revtex4-2}
\usepackage{amssymb,amsthm,amsmath,amsfonts}
\usepackage{graphicx,ulem,enumerate,mathptmx}
\usepackage[pdftex,dvipsnames,usenames]{xcolor}
\usepackage[colorlinks=true,urlcolor=blue,citecolor=blue,linkcolor=blue]{hyperref}

\begin{document}

\title{Driven Gaussian quantum walks}

\author{Philip Held}
	\affiliation{Integrated Quantum Optics Group, Institute for Photonic Quantum Systems (PhoQS), Paderborn University, Warburger Stra\ss{}e 100, 33098 Paderborn, Germany}
\author{Melanie Engelkemeier}
	\affiliation{Integrated Quantum Optics Group, Institute for Photonic Quantum Systems (PhoQS), Paderborn University, Warburger Stra\ss{}e 100, 33098 Paderborn, Germany}
\author{Syamsundar De}
	\affiliation{Integrated Quantum Optics Group, Institute for Photonic Quantum Systems (PhoQS), Paderborn University, Warburger Stra\ss{}e 100, 33098 Paderborn, Germany}
\author{Sonja Barkhofen}
	\affiliation{Integrated Quantum Optics Group, Institute for Photonic Quantum Systems (PhoQS), Paderborn University, Warburger Stra\ss{}e 100, 33098 Paderborn, Germany}
\author{Jan Sperling}
	\email{jan.sperling@upb.de}
	\affiliation{Theoretical Quantum Science, Institute for Photonic Quantum Systems (PhoQS), Paderborn University, Warburger Stra\ss{}e 100, 33098 Paderborn, Germany}
\author{Christine Silberhorn}
	\affiliation{Integrated Quantum Optics Group, Institute for Photonic Quantum Systems (PhoQS), Paderborn University, Warburger Stra\ss{}e 100, 33098 Paderborn, Germany}

\date{\today}


\begin{abstract}
	Quantum walks function as essential means to implement quantum simulators, allowing one to study complex and often directly inaccessible quantum processes in controllable systems.
	In this contribution, the notion of a driven Gaussian quantum walk is introduced.
	In contrast to typically considered quantum walks in optical settings, we describe the operation of the walk in terms of a nonlinear map rather than a unitary operation, e.g., by replacing a beam-splitter-type coin with a two-mode squeezer, being a process that is controlled and driven by a pump field.
	This opens previously unattainable possibilities for quantum walks that include nonlinear elements as core components of their operation, vastly extending their range of applications.
	A full framework for driven Gaussian quantum walks is developed, including methods to dynamically characterize nonlinear, quantum, and quantum-nonlinear effects.
	Moreover, driven Gaussian quantum walks are compared with their classically interfering and linear counterparts, which are based on classical coherence of light rather than quantum superpositions.
	In particular, the generation and boost of highly multimode entanglement, squeezing, and other quantum effects are studied over the duration of the nonlinear walk.
	Importantly, we prove the quantumness of the evolution itself, regardless of the input state.
	A scheme for an experimental realization is proposed.
	Furthermore, nonlinear properties of driven Gaussian quantum walks are explored, such as amplification that leads to an ever increasing number of correlated quantum particles, constituting a source of new walkers during the walk.
	Therefore, a concept for quantum walks is proposed that leads to---and even produces---directly accessible quantum phenomena, and that renders the quantum simulation of nonlinear processes possible.
\end{abstract}

\maketitle


\section{Introduction}

	For more than a century, quantum physics has reshaped our understanding of nature.
	Today, this theory informs previously inconceivable advancements of science and technology, such as quantum information theory that is based on superpositions of quantum bits rather than employing classical bits \cite{NC00}.
	Equivalent to the notion of a universal quantum computer is the concept of a quantum simulator \cite{F82} that is able to model any quantum processes, including quantum computations \cite{C09,LCETK10}.
	To realize such a desired simulation device, random walks that model classical processes have been elevated to the notion of quantum walks \cite{ADZ93,K03,V12}, again overcoming the restrictions of their classical counterparts.

	In optics, quantum walks are commonly implemented using photons and coherent light as the walker and linear optical elements, such as beam splitters (BSs), as coin-toss and step operations to realize the walk \cite{BMKSW99,MW14}.
	Regardless of the input state, the underlying network, being the core of any simulator platform, is typically confined to linear optics which cannot increase the quantum properties of the system;
	this is because this class of optical elements relies only on operations which are considered as classical mode transformations from the perspective of quantum optics \cite{KSBK02,X02,VS14}, unable to exceed classical interference phenomena \cite{KRS03,HBF03,JPK04,FIPL06}.
	More generally speaking, even today, it is hard to draw a clear line between analog, yet classical computers---exploiting classical interference---and true quantum simulators \cite{D20,Aetal21}.

	Despite this hurdle, remarkable experimental and theoretical progress has been made since early proofs-of-concept experiments of quantum walks \cite{DLXSWZH03}.
	For example, different transport phenomena, including disorder and various diffusion regimes, have been simulated through optical quantum walk implementations \cite{BFLKAW10,SCPGJS11,GLBSFFSCM19,GDLBSMS21}.
	Larger coin spaces and time-dependent coins nullify restriction to static binary heads-or-tails values to determine the walkers propagation direction \cite{SCPGMAJS10,LMNPGBJS19}.
	Different geometries on which the walker can evolve have been realized, including walks on circles with periodic boundary conditions \cite{LMNPGBJS19,MPPMPIWOT13,BLZTX17}.
	Quantum correlations in various quantum walk scenarios have led to a much deeper understanding for how quantum features can propagate, with entanglement being the most prominent and most useful quantum correlation property \cite{GMPKDNSACS13,CORGFSNSM13,MLP13,RLPPSS16,GLN19,Wetal20}.
	Other successful applications include studying topological phases \cite{A12}, measurement-induced coherence effects \cite{NBKSSGPKJS18}, phase-space-based characterizations \cite{SMSGEHS09}, the ability to dynamically create and annihilate photons (i.e., walkers) \cite{HBSJS16}, nonlocalized input states \cite{Setal19}, large-scale fiber-assisted quantum walks \cite{BFBFKKW16}, single- and multiphoton interference \cite{PA07,BLMS09,NDBMTSGJS20}, etc.
	These examples showcase the overall success of quantum walks as a versatile tool in theoretical and experimental physics, in general, and in optical implementations, in particular.

	Additionally and gradually, the untapped potential of nonlinear components has been recognized as a means to advance quantum walks even farther where already weak nonlinearities can have a significant impact \cite{PS08,VFF19}.
	Such approaches, for example, place the entire walk in a nonlinear medium \cite{NPR07}, consider enhancement through externally driven cavities \cite{AP05}, and exploit feed-forward mechanisms \cite{SWH14}.
	Other recent approaches lay the fundamentals to embed quantum walks in nonlinear waveguide arrays \cite{Setal14,BXLZZ16}.
	However, owing to the inherent complexity of nonlinear systems, what is common to most proposals is that a comprehensive framework is missing, including a way to predict and quantify the emerging quantum features in such nonlinear simulators.
	Also, nonlinearities are often considered as an addition---and not a key functional component---of otherwise linear walks.

	An essential, nonlinear component in quantum optics is the squeezing operation, leading to the notion of squeezed light \cite{W83}.
	This operation is, for example, achieved through parametric down-conversion (PDC) \cite{WKHW86}.
	See, e.g., Refs. \cite{L00,VW06,A12book} for thorough introductions.
	Since squeezed light's defining feature is a field fluctuation below vacuum fluctuations, squeezed states are not only highly relevant from a fundamental point of view but also from a practical perspective \cite{WPGCRSL12,ARL14,S17}.
	For instance, in metrology, such states allow for a sensitivity that exceeds classical boundaries, e.g., finding applications in gravitational wave detectors to monitor minuscule spacetime perturbations \cite{C81,GDDSSV13,LIGO11}.
	Quantum communications also benefit from squeezed light as a means to establish quantum key distribution \cite{R99,H00,CLA01}.
	Proof-of-concept protocols that outperform any classical simulations, such as boson sampling \cite{AA13}, can be realized using squeezed states as well \cite{HKSBSJ17,Zetal20}.
	This method can be further enhanced by dynamically producing photons within the boson sampler \cite{BBSKHJS17}, dubbed driven Gaussian boson sampling.
	In multimode scenarios, squeezed light provides an excellent basis to realize and study complex forms of multipartite entanglement \cite{CMP14,GSVCRTF15}, exceeding bipartite quantum correlations \cite{GSVCRTF16}.
	See also Ref. \cite{FT20} for a detailed review about multimode squeezed states in theory and practice.
	For the aforementioned reasons, squeezed states play a key role for continuous-variable quantum information processing \cite{ARL14,WPGCRSL12}.
	Consequently, for our purpose, PDC can be considered as an ideal, first candidate for a nonlinear operation to significantly extend linear quantum walks.

	In this contribution, we establish and comprehensively investigate the notion of a driven Gaussian quantum walk (DGQW).
	By replacing ordinary BSs with PDC-type operations, we advance the function of quantum walk to the nonlinear regime.
	It is shown that prominent quantum features, such as entanglement, squeezing, and photon-number correlations, are introduced and amplified through this added functionality itself, without requiring sophisticated, nonclassical input states.
	Our DGQW architecture is compared with its linear counterparts, demonstrating a superior quantum performance.
	A full framework for DGQWs is developed, which applies to nonlinear coin and step operations for arbitrary coin dimensionalities and geometries on which the walker can propagate.
	To exemplify the benefits of DGQWs, a comprehensive and exact analysis of the evolution of a quantum walker on a line with a biased coin and periodic boundary conditions is carried out.
	Measurable means to quantify the quantum effects in DGWQs are formulated.
	Our research further includes the dynamical study of nonlinear features, such as amplification and unavoidable noise contributions.
	Generalizations of our approach are briefly discussed and a proposal for an experimental implementation is outlined.
	Therefore, a theoretical method is put forward that is based on existing experimental techniques and leads to interesting, previously unexplored quantum effects in the nonlinear evolution of quantum walks with practical relevance for future quantum simulator platforms.


\section{Preliminaries and extended motivation}
\label{Sec:Motivation}

	In this section, we firstly introduce a fine-grained separation of walk-based simulators.
	Specifically, not only do we distinguish between random walks and quantum walks, but we also introduce the notion of a coherent walk as an intermediate stage, which is based on general interference;
	i.e., a coherent walk can be realized, for example, within classical wave theories, too.
	Secondly, we then consider what happens when the linear coin operation that describes the walk is replaced by a distinctively different nonlinear coin, foreshadowing nonlinear, quantum, and quantum-nonlinear effects of the evolution we comprehensively study in the following sections.

\subsection{Coherent walks: The different origins of interference}

\begin{figure}
	\includegraphics[width=\columnwidth]{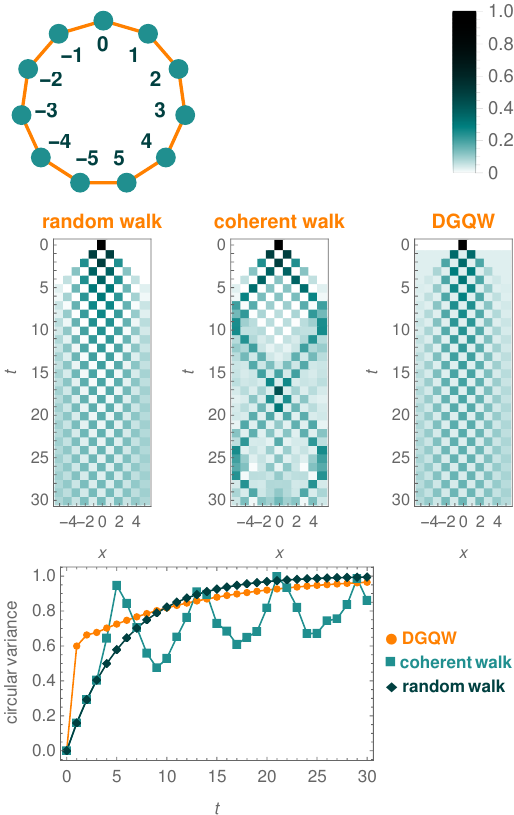}
	\caption{%
		A random walk on a graph (top, left) with $d=11$ vertices---position labels $x\in\{-5,\ldots,5\}\,\mathrm{mod}\,d$---is carried out.
		The output probability $P(x)$ (middle, left) for a time $t$ is encoded in the depicted brightness (top, right).
		Since $d$ is odd, even and odd positions can gain a nonzero probability after five steps when the walker, who is initialized at $x=0$, crosses the periodic boundaries at $x=\pm 5\,\mathrm{mod}\,d$.
		The circular variance (bottom) as a measure of random walk's spread increases with the number of steps and eventually saturates at the maximum value one.
		The position probabilities of a quantum walk (middle, center) with a single-photon input and a polarization-based Hadamard coin on the same graph is depicted.
		This walk is identical to the intensity distribution---always normalized to the total intensity---of a coherent walk, e.g., with coherent input amplitudes $\alpha_{(x=0,c=+)}=\sqrt{10}=i\alpha_{(x=0,c=-)}$ and $\alpha_{(x,c)}=0$ otherwise;
		thus, on average $\sum_{x,c}|\alpha_{x,c}|^2=20$ photons propagate in the optical network.
		The circular variance (bottom) oscillates with $t$ for the coherent walk (likewise, the single-photon walk), clearly demonstrating interference.
		Last, a DGQW (middle, right) is shown with a nonlinear coin that is a seeded PDC process with infinite squeezing and with the same input as the coherent walk.
		While the initial step increases the circular variance (bottom) rather significantly, subsequent steps show a smaller increment than classical random walks, but without exhibiting signs of oscillations like coherent walks.
		Yet, we claim---and are going to prove---that the DGQW exhibits stronger quantum superposition features than the coherent walk.
	}\label{fig:motivation}
\end{figure}

	Random walks present a uniquely useful tool for simulating classical transport phenomena.
	As one example, we may explore the random motion on a looped linear chain with $d$ nodes; see Fig. \ref{fig:motivation} (top, left).
	A walker that is initialized at position $x=0$ tosses a fair coin and determines whether to make a step in a clockwise (coin value $c=+$) or counterclockwise ($c=-$) direction.
	The question now is where the walker is found after $t$ iterations of coin tosses and steps.
	When averaging over multiple runs, this yields a probability distribution $P(x)$ over the position $x$ which eventually covers all vertices of the graph.
	To quantify the spread on a periodic geometry, one can determine the circular variance \cite{F93}, which reads $1-|\sum_{x=0}^{d-1}P(x)e^{2\pi i x/d}|$ in our case.
	In the asymptotic limit, $t\to\infty$, the random walker is uniformly spread over all nodes of the looped graph (Fig. \ref{fig:motivation}, bottom), resulting in the maximal value one for the circular variance as this walk describes a diffusive transport scenario.

	While the classical walker can be treated as a particle, quantum walks interpret the walker as a quantum particle, including its ability to interfere \cite{K03,V12}; see middle, center plot in Fig. \ref{fig:motivation}.
	Thus, a quantum walk is now able to simultaneously explore the whole---often rather complex---geometry, rather than requiring multiple runs as in incoherent random walk.
	This feature of being in a superposition of all positions $x$ of a network is central for the function of many quantum simulation approaches \cite{KRS03,HBF03}.

	Nonetheless, common experiments use linear optics and classical coherent light alone to achieve many results that were initially described in a quantum walk framework \cite{BMKSW99,MW14,JPK04,FIPL06}, such as Anderson localization \cite{SCPGJS11} and faster spreads than possible for random walks \cite{SCPGMAJS10}.
	Clearly, classical light can mimic coherent interference effects because of its intrinsic wave properties.
	In particular, this can be achieved by replacing the probability amplitudes of quantum states with the multimode coherent field amplitudes of light, $\alpha_{x,c}$.
	Equivalently, the probability distribution is then represented by the photon-number (or intensity) distribution over the positions $x$, $P(x)\propto |\alpha_{x,+}|^2+|\alpha_{x,-}|^2$, which is normalized to the total photon number (intensity).
	See Fig. \ref{fig:motivation} for a Hadamard walk that utilizes an unbiased coin that reflects and transmits $50\%$ of the intensity and utilizing only coherent light.

	To distinguish quantum walks that truly require quantum physics from scenarios in which any form of interference suffices, we here refer to the latter as coherent walks and to the former as genuine quantum walks.
	Moreover, even if we have a genuine quantum walk, it is often unclear if an observed interference effect is a result of the quantumness of process itself or if it is merely a transformation of the nonclassical properties that were already present in the initial state, which applies to, e.g., linear single-photon walks \cite{Setal19,BFLKAW10}, two-photon interferences \cite{NDBMTSGJS20}, and even other interferometric schemes \cite{RSG19}.

	An example of a genuine quantum walk that produces quantumness via the walk is the DGQW, which is also depicted in Fig. \ref{fig:motivation} (middle, right).
	However, quantum effects do not easily manifest themselves in the photon-number distribution;
	in fact, when focusing on the depiction in Fig. \ref{fig:motivation} alone, a DGQW seems to be more akin to an incoherent random walk.
	Thus, higher-order quantum coherence effects are to be investigated to prove the buildup of quantum phenomena.

	In the remainder of this work, we address all points outlined above.
	For this purpose, we establish the notion of a DGQW.
	In a quantum-optical setting, this genuine quantum walk uses a pump field to drive a nonlinear process that produces nonlocal quantum superposition.
	Furthermore, we prove that the evolution itself causes the quantum phenomena.
	To this end, a rigorous study of quantum effects in quantum walks is carried out by formulating higher-order correlation functions and experimentally accessible (co-)variance-based quantumness metrics.
	Thereby, DGQWs provide a conceptual revision of how we characterize walk processes in terms of the resulting classical and quantum coherence properties and further separating dynamic phenomena from nonclassical input properties.

\subsection{A nonlinear coin: From classical to quantum interference}

	Another classification of simulations can be made in terms of the kind of evolution.
	For example, the usual optical version of a coherent walk employs linear optics.
	Here, however, we exploit nonlinear components, thus driving a nonlinear quantum walk rather than a linear walk.
	For sake of exposition, we focus on nonlinear coin operations for the time being, and the possibility of nonlinear step operations is embedded in the general framework as presented in Sec. \ref{Sec:GenModel}.
	In general, the transition from a linear to nonlinear interferometry offers coherence features which are distinctively different from their linear counterparts \cite{LSSSMFSHS21}.
	Please note that we use the word coin as a shorthand for the coin operation, being different from the quantum state of the coin, the coin state.

\begin{figure}
	\includegraphics[width=\columnwidth]{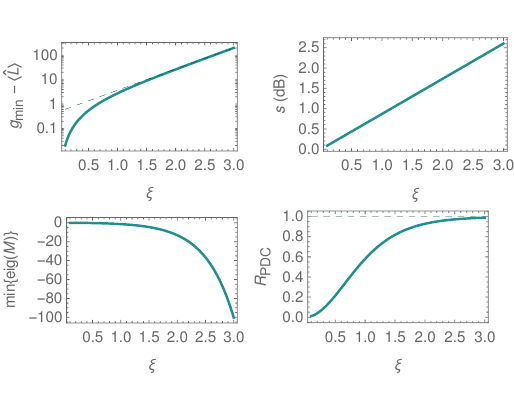}
	\caption{%
		Selected properties of the nonlinear coin operation.
		The top-left plot shows the entanglement between polarizations that is generated as a function of the squeezing parameter $\xi$ (Sec. \ref{Subsec:Entanglement}).
		The dashed line depicts the asymptotic behavior.
		The squeezing in decibel in the top-right plot is, of course, proportional to the squeezing parameter, with a proportionality constant $2\log_{10}e\approx 0.8686$ (Sec. \ref{Subsec:Squeezing}).
		Nonclassical photon-number correlations between the output photons in both polarizations are shown in the bottom-left plot in terms of negativities (Sec. \ref{Subsec:PhotNoCorr}).
		While the other properties considered here are invariant under the coherent amplitudes of the input, particle correlations are not and are specifically shown for a vacuum input, demonstrating that photon-pair---and correlations between them---are produced via the PDC coin.
		Finally, the bottom-right plot shows the splitting ratio between transmitted and reflected polatizations $R_\mathrm{PDC}$ [Eq. \eqref{eq:SplittingRatio}].
		A value $R_\mathrm{PDC}=1$ is approached for $\xi\to\infty$ (dashed line), resembling a fair $50:50$ coin.
	}\label{fig:PDCCoin}
\end{figure}

	To highlight the advancements that can be achieved by transitioning from a linear to a nonlinear regime, we compare a single coin operation for linear and nonlinear quantum walks.
	In linear optics, all BS-coin maps can be implemented by waveplates and polarizing BSs when, as we here do, the coin acts on the polarization of light.
	Then, for example, the horizontal and vertical field components can be identified with the two coin values $c=+$ and $c=-$, respectively.
	The polarization determines the propagation direction for the subsequent shift operation, implemented by a polarizing BS.

	For our nonlinear coin, we consider a stimulated nonlinear process, PDC, in which photon pairs are coherently added to the incident light, the so-called seed of that process.
	For the example of vacuum input, this results in the two-mode squeezed vacuum, $\sqrt{1-|\lambda|^2}\sum_{n=0}^\infty\lambda^n|n\rangle\otimes|n\rangle$ ($|\lambda|=\tanh|\xi|$), which is entangled and includes the same number of photons $n$ in both output polarizations.
	The PDC process serves as the counterpoint to the BS-based coherent quantum walk approach in this work.
	The quantum interference characteristics of a single PDC coin are discussed in the following, serving as a precursor to a full and general DGQW description in the next sections.
	See Fig. \ref{fig:PDCCoin} for a summary.

	In the linear scenario of a BS coin, we map the bosonic field operators for the coin states $+$ and $-$ as $\hat a_+\mapsto\tau\hat a_++\rho \hat a_-$ and $\hat a_-\mapsto -\rho^\ast\hat a_++\tau^\ast\hat a_-$, respectively, with $|\tau|^2+|\rho|^2=1$.
	The output intensity, represented by the photon number, is thus obtained through
	\begin{equation}
		\label{eq:BSone}
	\begin{aligned}
		\langle\hat a_\pm^\dag\hat a_\pm\rangle
		\mapsto
		&|\tau|^2\langle\hat a_\pm^\dag\hat a_\pm\rangle
		+|\rho|^2\langle\hat a_\mp^\dag\hat a_\mp\rangle
		\\
		&\pm\rho^\ast\tau\langle\hat a_-^\dag\hat a_+\rangle
		\pm\tau^\ast\rho\langle\hat a_+^\dag\hat a_-\rangle.
	\end{aligned}
	\end{equation}
	Therein, the first and second terms, respectively, describe the transmitted ($\propto |\tau|^2$) and reflected ($\propto |\rho|^2$) contributions of the incident light to the output intensity.
	The last two terms describe the interference of the involved quantum fields.

	By contrast, the transformation $\hat a_\pm\mapsto\mu\hat a_\pm+\nu\hat a^\dag_\mp$, with $|\mu|^2-|\nu|^2=1$, captures a two-mode PDC process \cite{L00,A12book,VW06}, which depends on the field operator and its Hermitian conjugate.
	This yields the following output photon numbers:
	\begin{equation}
		\label{eq:PDCone}
	\begin{aligned}
		\langle\hat a_\pm^\dag\hat a_\pm\rangle
		\mapsto
		&|\mu|^2\langle\hat a_\pm^\dag\hat a_\pm\rangle
		+|\nu|^2\langle\hat a_\mp^\dag\hat a_\mp\rangle+|\nu|^2
		\\
		&+\nu^\ast\mu\langle\hat a_\pm\hat a_\mp\rangle
		+\mu^\ast\nu\langle\hat a_\pm^\dag\hat a_\mp^\dag\rangle,
	\end{aligned}
	\end{equation}
	where we used the fundamental commutation relation $\hat a_\mp\hat a_\mp^\dag=\hat a_\mp^\dag\hat a_\mp+\hat 1$ in the expansion.
	Please recall that $\hat a\mapsto\hat a^\dag$ is not a linear map;
	specifically, it does not satisfy homogeneity, $(c\hat a)^\dag=c^\ast\hat a^\dag\neq c\hat a^\dag$ for complex numbers with $\mathrm{Im}(c)\neq0$.
	Similarly to Eq. \eqref{eq:BSone}, the first two terms in Eq. \eqref{eq:PDCone} resemble transmitted ($\propto|\mu|^2$) and reflected ($\propto|\nu|^2$) intensities, respectively, and the last two terms constitute the interference.

	We also have a number of profound dissimilarities between PDC and BS coins.
	Firstly, $|\mu|^2\geq1$ implies that the input light is amplified upon transmission, which is one quantum-physical interpretation of the PDC process under study that is enabled by driving this nonlinear process by a pump pulse.
	(Recall that $|\tau|^2\leq 1$ holds true for BS-type transmitivity.)
	Secondly, in the quantum domain, such an amplification is further connected to a certain amount of excess noise, attributed to the constant third term $|\nu|^2$ in Eq. \eqref{eq:PDCone} that is not present in the linear case;
	see, e.g., Ref. \cite{ELDBDPSS20} for an in-depth analysis of the interplay of amplification, loss, and excess noise.
	Commonly, one understands this noise as the result of the amplification of vacuum fluctuations of the input.

	Thirdly, the interference in the nonlinear case is determined through $\langle \hat a_+\hat a_-\rangle$ and $\langle \hat a_+^\dag\hat a_-^\dag\rangle=\langle \hat a_+\hat a_-\rangle^\ast$, contrasting the linear scenario with $\langle \hat a_+^\dag\hat a_-\rangle$ and $\langle \hat a_-^\dag\hat a_+\rangle=\langle \hat a_+^\dag\hat a_-\rangle^\ast$.
	In particular, those terms lead to different phase relations between the output light fields since each scenario depends on a different combination of annihilation ($\hat a_\pm$) and creation ($\hat a_\pm^\dag$) operators.
	Also, the interference terms for the nonlinear case in Eq. \eqref{eq:PDCone} are preceded by like signs while opposite signs for the $+$ and $-$ polarizations occur in Eq. \eqref{eq:BSone}.
	This results in a different interplay of constructive and destructive interference;
	compare, for instance, the interference patterns in the center and right plot in the middle row of Fig. \ref{fig:motivation}.
	Fourthly, various quantum properties between the output polatizations are generated by a PDC coin but not by a BS coin;
	see first three plots in Fig. \ref{fig:PDCCoin}.

	Finally, we can relate the transmitted and reflected photon numbers for each scenario, Eqs. \eqref{eq:BSone} and \eqref{eq:PDCone}.
	Specifically, the respective splitting ratios are
	\begin{equation}
	\label{eq:SplittingRatio}
		R_\mathrm{BS}=\frac{|\rho|^2}{|\tau|^2}=\tan^2\vartheta
		\quad\text{and}\quad
		R_\mathrm{PDC}=\frac{|\nu|^2}{|\mu|^2}=\tanh^2\xi,
	\end{equation}
	which means $\tau\propto\cos\vartheta$ and $\rho\propto\sin\vartheta$ as well as $\mu\propto \cosh\xi$ and $\nu\propto \sinh\xi$ hold true, while ignoring phases.
	Clearly, $\vartheta$ defines a coin's rotation angle of the BS, thus $0\leq R_\mathrm{BS}<\infty$.
	The angle $\theta=\pi/4$, for example, yields a fair coin, $R_\mathrm{BS}=50\%/50\%=1$.
	For the PDC, $\xi$ is the dubbed the squeezing parameter and determines the coin's action.
	In this nonlinear scenario, we have $0\leq R_\mathrm{PDC}\leq 1$, where the upper bound is approached for an infinite squeezing, $\xi\to\infty$.
	We can also equate both splitting ratios, $R_\mathrm{BS}=R_\mathrm{PDC}$, allowing for a comparison of both types of quantum walks with identical splitting ratios.
	This applies as long as $R_\mathrm{BS}< 1$ is obeyed.
	The bottom-right graph in Fig. \ref{fig:PDCCoin} shows the splitting ratios of the PDC coin as a function of the squeezing parameter $\xi$.
	The asymptotic case $\xi\to\infty$ can be established by introducing a renormalization, $\mu'=\mu/|\mu|$ and $\nu'=\nu/|\mu|$, such that $|\mu'|^2=1\stackrel{\xi\to\infty}{\to}1$ and $|\nu'|^2=\tanh^2\xi\stackrel{\xi\to\infty}{\to}1$ remain bounded quantities in the infinite-squeezing limit.
	This renormalization was used for the DGQW in Fig. \ref{fig:motivation} to ensure a $50:50$ PDC coin that is compared with a fair BS coin for the coherent Hadamard walk that utilizes a BS operation, $R_\mathrm{PDC}=1=R_\mathrm{BS}$.

	To summarize, BS and PDC coins are rather distinct concepts when it comes to their interference characteristics, being essential for implementing quantum simulator platforms.
	Specifically, BS operations cannot produce nonclassical outputs \cite{VS14}, but PDC coins intrinsically generate seminal quantum effects, such as entanglement.
	In the general framework we consider next, Secs. \ref{Sec:GenModel} and \ref{Sec:Evolution}, we show that the linear case may be embedded in the nonlinear one, indeed.
	Consequently, nonlinear interference can include and even exceed linear ones, allowing one to simulated quantum-physical processes that are inaccessible in the linear regime.
	Furthermore, we are going to comprehensively show and analyze the distinct differences between entire quantum walks using BS-type operations and PDC-like processes that are inspired by the initial considerations conducted here for a single coin operation.


\section{General model for DGQWs}
\label{Sec:GenModel}

	After studying the impact of one nonlinear coin, we now elaborate on the general theoretical description of DGQWs.
	This includes scenarios in which one can have PDC-type and BS-type coins at the same time, the possibility of nonlinear step operations, embedding single-mode squeezers within the walk, using PDC sources to produce general Gaussian input states, etc.
	Furthermore, the methodology presented here applies to arbitrarily dimensional position spaces and geometries, as well as coins that can take more than two values.

	For achieving the sought-after broadly applicable model, it is convenient to introduce vectors of annihilation and creation operators, $\vec{\hat a}=[\hat a_i]_i$ and $\vec{\hat a}^\dag=[\hat a_i^\dag]_i$.
	(We remark that those tuples of linear matrices can be considered as elements of a module, i.e., loosely speaking, a vector field based on a ring.)
	The index $i$ is a pair, $i=(x,c)$, where $x$ encodes the possibly high-dimensional position, e.g., on a complex graph structure, and $c$ assigns one of various coin values.
	The propagation of the system after $t$ steps shall be described by a matrix $T$, which can combine annihilation and creation operators,
	\begin{equation}
	\label{eq:FundamentalInOut}
		\begin{bmatrix}
			\vec{\hat a} \\ \vec{\hat a}^\dag
		\end{bmatrix}
		\mapsto 
		\begin{bmatrix}
			\vec{\hat b} \\ \vec{\hat b}^\dag
		\end{bmatrix}
		=\underbrace{
			\begin{bmatrix}
				U & V \\ V^\ast & U^\ast
			\end{bmatrix}
		}_{=T}
		\begin{bmatrix}
			\vec{\hat a} \\ \vec{\hat a}^\dag
		\end{bmatrix},
	\end{equation}
	where $U$ and $V$ are matrices and ``$b$'' denotes bosonic operators of output modes.
	Note that second line of the input-output relation \eqref{eq:FundamentalInOut} is obtained by Hermitian conjugation of the first line, $(\hat b_i)^\dag$, and the nonlinear character is encoded via $V\neq0$, mixing annihilation and creation operators.
	Of course, the output shall obey fundamental commutation relations, $[\hat b_{i},\hat b_{i'}]=0$ and $[\hat b_{i},\hat b_{i'}^\dag]=\delta_{i=i'}\hat 1$, where the Kronecker symbol takes the values $\delta_\mathrm{true}=1$ and $\delta_\mathrm{false}=0$.
	This is satisfied if and only if
	\begin{equation}
	\label{eq:UVconditions}
		UV^\mathrm{T}=VU^\mathrm{T}
		\quad\text{and}\quad
		UU^{\ast\mathrm{T}}-VV^{\ast\mathrm{T}}=\mathrm{id}
	\end{equation}
	are obeyed, where ``$\mathrm{id}$'' denotes the identity matrix, being different from the identity operator $\hat 1$.
	Similarly, for the sake of differentiation, we write ``$\ast\mathrm{T}$'' to indicate the Hermitian conjugate of a matrix, rather than using $\dag$ that, throughout this work, is exclusively applied to quantum-physical operators.
	It is also worth mentioning that the inverse of $T$ is given by
	\begin{equation}
	\label{eq:InverseTrafo}
		T^{-1}=\begin{bmatrix}
			U^{\ast\mathrm{T}} & -V^\mathrm{T}
			\\
			-V^{\ast\mathrm{T}} & U^\mathrm{T}
		\end{bmatrix},
	\end{equation}
	allowing us to conversely describe input fields through output modes and which follows from Eq. \eqref{eq:UVconditions}.

	It is easy to see [e.g., via Eq. \eqref{eq:UVconditions}] that $V=0$ implies a purely unitary---thus, linear---evolution.
	For $V\neq 0$, we mix the field and its Hermitian conjugate field, defying homogeneity of a linear map as mentioned earlier.
	Thus, our general DGQW model does fully include linear quantum walks as a specific instance but also farther extends to nonlinear scenarios.

	Moreover, the above class of operations $T$ are so-called Gaussian operations \cite{WPGCRSL12,ARL14}---further supplemented by displacement operations that are not focused on here.
	In particular, such operations map any Gaussian input state to a Gaussian output state.
	In this work, we mostly consider Gaussian inputs, specifically classical and factorizable input light that is described via multimode coherent states,
	\begin{equation}
	\label{eq:InputState}
		|\vec\alpha\rangle=\bigotimes_i|\alpha_i\rangle,
	\end{equation}
	with arbitrary field amplitudes $\alpha_i$ for the $i$th input mode.
	Because of this choice of initial states, genuine quantum effects at the output, such as squeezing and entanglement, must be a result of the nonlinear process under consideration;
	see Fig. \ref{fig:PDCCoin} for an example.
	In fact, it is known that unitary processes alone ($V=0$) are not capable of producing such quantum phenomena for these kinds of inputs \cite{KSBK02,X02,VS14}.

	In the continuation of this section, we consider how different measurable quantities propagate under the general evolution in Eq. \eqref{eq:FundamentalInOut}, whose details are provided in several appendices.
	Thus, we here discuss the results and refer to the corresponding technical calculation in the relevant appendix where and when necessary.
	Furthermore, we here describe how we can infer nonlinear and quantum and quantum-nonlinear features from those measurable quantities for the general case of DGQWs that harness the potential of Gaussian operations beyond BS-like unitary maps.

\subsection{Covariance matrix formalism}
\label{Subsec:Squeezing}

	Gaussian states are wholly described by their first- and second-order moments \cite{WPGCRSL12,ARL14}.
	We are going to analyze the propagation of those quantities for general DGQWs.
	Before that, we make a simple observation for the quantum-physical expectation values of bosonic field operators,
	\begin{equation}
		\label{eq:AmplitudePropagation}
		\vec\alpha=\langle\vec{\hat a}\rangle
		\,\,\mapsto\,\,
		\langle\vec{\hat b}\rangle
		=U\langle\vec{\hat a}\rangle
		+V\langle\vec{\hat a}^\dag\rangle
		=U\vec\alpha+V\vec\alpha^\ast
		=\vec \beta,
	\end{equation}
	which is a straightforward consequence of Eq. \eqref{eq:FundamentalInOut} telling us that the coherent amplitudes transform accordingly.

	The real-valued mean field and its conjugate momentum can be measured, for example, via balanced homodyne detection \cite{VW06}.
	The corresponding quadrature and momentum operators are
	\begin{equation}
		\label{eq:AmplitudesTransform}
		\vec{\hat q}=\frac{\vec{\hat b}+\vec{\hat b}^\dag}{\sqrt{2}}
		\quad\text{and}\quad
		\vec{\hat p}=\frac{\vec{\hat b}-\vec{\hat b}^\dag}{i\sqrt{2}},
	\end{equation}
	jointly describing all output fields.
	The combination of quadratures and momenta yields the displacement vector, which can be obtained from the input displacement as
	\begin{equation}
		\begin{bmatrix}
			\langle\vec{\hat q}\rangle \\ \langle\vec{\hat p}\rangle
		\end{bmatrix}
		{=}\overbrace{\begin{bmatrix}
			\mathrm{Re}(U)+\mathrm{Re}(V) & -\mathrm{Im}(U)+\mathrm{Im}(V)
			\\
			\mathrm{Im}(U)+\mathrm{Im}(V) & \mathrm{Re}(U)-\mathrm{Re}(V)
		\end{bmatrix}}^{=T^\prime}
		\begin{bmatrix}
			\sqrt{2}\mathrm{Re}(\vec\alpha) \\ \sqrt{2}\mathrm{Im}(\vec\alpha)
		\end{bmatrix}\!,
	\end{equation}
	generally constituting a symplectic transformation $T'$, simplifying to a rotation for $V=0$.

	Beyond those first-order moments, the covariances between quadratures and/or momenta describe the quantum-noise properties.
	The variances are described through expectation values of the form $\langle(\Delta\hat q)^2\rangle$, where $\Delta\hat q=\hat q-\langle\hat q\rangle\hat 1$, which similarly applies to momentum-type observables, $\Delta\hat p$.
	Because of the noncommuting nature of the field operators, $[\hat q,\hat p]=i\hat 1\neq0$, covariances are described in a symmetrical manner via the anticommutator \cite{WPGCRSL12,ARL14}, e.g., $\langle\{\Delta\hat q,\Delta\hat p\}\rangle/2=(\langle \hat q\hat p\rangle+\langle \hat p\hat q\rangle)/2-\langle\hat q\rangle\langle\hat p\rangle$.
	Variances and covariances are then collected in the covariance matrix, $C$.
	For coherent light, such as our input in Eq. \eqref{eq:InputState}, the covariance matrix is the identity scaled by one half.
	By applying the detailed calculation in Appendix \ref{App:MomentTransform}, we further obtain the output covariance matrix
	\begin{equation}
	\begin{aligned}
		C=&T^\prime\frac{1}{2}\begin{bmatrix}
			\mathrm{id} & 0 \\ 0 & \mathrm{id}
		\end{bmatrix}T^{\prime\mathrm{T}}
		\\
		=&\frac{1}{2}\begin{bmatrix}
			\mathrm{id} & 0 \\ 0 & \mathrm{id}
		\end{bmatrix}
		+\begin{bmatrix}
			\mathrm{Re}(UV^\mathrm{T})
			& \mathrm{Im}(UV^\mathrm{T})
			\\
			\mathrm{Im}(UV^\mathrm{T})
			&
			-\mathrm{Re}(UV^\mathrm{T})
		\end{bmatrix}
		\\
		&+\begin{bmatrix}
			\mathrm{Re}(VV^{\ast\mathrm{T}})
			& -\mathrm{Im}(VV^{\ast\mathrm{T}})
			\\
			\mathrm{Im}(VV^{\ast\mathrm{T}})
			&
			\mathrm{Re}(VV^{\ast\mathrm{T}})
		\end{bmatrix}.
	\end{aligned}
	\end{equation}
	The latter two contributions---additional terms when compared with the input---vanish for a purely linear quantum walk, $V=0$.
	However, this part is essential to characterize the buildup of the quantum phenomenon that is squeezing \cite{W83}.

	Squeezing describes the reduction of quantum noise in certain quadratures and momenta below the value for vacuum---or classical coherent states that exhibit the same amount of noise.
	The multimode covariance matrix $C$ of our DGQW yields the amount of multimode squeezing through its minimal eigenvalue, $\min\{\mathrm{eig}(C)\}$.
	That is, whenever this eigenvalue is below the vacuum level $1/2$, squeezing is verified.
	Commonly, the level of squeezing relative to vacuum is expressed in decibel,
	\begin{equation}
		s\,\mathrm{(dB)}=-\log_{10}\left(\frac{\min\{\mathrm{eig}(C)\}}{1/2}\right),
	\end{equation}
	where a positive value certifies squeezing and a negative value describes anti-squeezing (i.e., a noise increment).
	It is additionally noteworthy that when restricting to certain submatrices of $C$, the minimal eigenvalue of this restricted covariance matrix allows one to determine the squeezing for a selection of modes, such as single-mode squeezing.
	For a single PDC coin, the squeezing was shown in Fig. \ref{fig:PDCCoin} (top, right).

\subsection{Particle-number-based characterization}
\label{Subsec:PhotNoCorr}

	Simpler than a balanced homodyne detection is a measurement of photon numbers, directly providing the statistics of the particle number for each mode.
	Random, coherent, and quantum walks basically assess the probability of the occurrence of the walker at a given node of a network through the (absolute, relative, or otherwise normalized) number of particles in each mode, which, for light, is proportional to the intensity.
	The corresponding photon-number operator is defined as $\hat n_i=\hat b_i^\dag\hat b_i$ and allows us to study the photon-number dynamics in our DGQWs.
	(Detailed calculations of the photon number and its correlations can be found in Appendix \ref{App:MomentTransform}.)

	The mean photon number for each mode can be combined in a vector, which is obtained from the input as
	\begin{equation}
		\langle\vec{\hat n}\rangle=\left[|\beta_i|^2+(VV^{\ast\mathrm{T}})_{i,i}\right]_i.
	\end{equation}
	using the values for $\vec\beta$ from Eq. \eqref{eq:AmplitudesTransform} and ``$(\cdots)_{i,i'}$'' identifying the corresponding entry of a matrix.
	Again, $V\neq 0$ has a significant impact on the photon numbers in all modes.
	In particular, this yields an extra noise contribution $(VV^{\ast\mathrm{T}})_{i,i}$, which is independent of the input light, as a result of the nonlinear operations under study.
	For example, we can quantify the overall excess noise by summing over the difference of the measured photon number and the coherent contribution, $\sum_i\left(\langle \hat n_i\rangle-|\beta_i|^2\right)=\mathrm{tr}(VV^{\ast\mathrm{T}})$, allowing us further to study the background level of this noise in the walker's evolution.
	While noise comes with a certain negative connotation, let us emphasize here that the addition to the photon number also means that new walkers in the form of quantum particles are generated during the walk.

	Typically, one focuses on the particle-number distribution to study transport phenomena in quantum walks, such as assessed by the position spread of this distribution over time.
	For certifying quantum phenomena, however, we can additionally correlate particle numbers, e.g., between different positions to infer a nonclassical behavior \cite{A88}, even beyond squeezing \cite{AT92}.
	To this end, one can utilize normally ordered moments of the experimentally determined photon-number distribution, $\langle{:} \hat n_i\hat n_{i'}{:}\rangle=\langle\hat n_i\hat n_{i'}\rangle-\delta_{i=i'}\langle \hat n_i\rangle=\langle \hat b_i^\dag\hat b_{i'}^\dag\hat b_{i}\hat b_{i'}\rangle$, being a factorial moment of the multimode photon-number distribution and a quantum-optical intensity correlation function \cite{VW06}.
	The central moments of such correlations form a matrix of moments, which is given from the input as
	\begin{equation}
	\begin{aligned}
		M=&\left[\langle{:} (\Delta\hat n_i)(\Delta\hat n_{i'}){:}\rangle\right]_{i,i'}
		\\
		=&[UV^\mathrm{T}]\circ[U V^{\mathrm{T}}]^\ast
		+[VV^{\ast\mathrm{T}}]\circ[V V^{\ast\mathrm{T}}]^\ast
		\\
		&+2\mathrm{Re}\left(
			[U V^\mathrm{T}]^\ast\circ[\vec\beta\vec\beta^\mathrm{T}]
		\right)
		+2\mathrm{Re}\left(
			[V V^{\ast\mathrm{T}}]\circ[\vec\beta^\ast\vec\beta^\mathrm{T}]
		\right),
	\end{aligned}
	\end{equation}
	where ``$\circ$'' denotes the Schur (i.e., element-wise) product of matrices;
	see Appendix \ref{App:MomentTransform} for the derivation.
	Whenever $M$ exhibits a negative eigenvalue, nonclassical photon-number correlations are identified \cite{VW06,A12book}, which can happen only if $V\neq0$ holds true.
	For example, in Fig. \ref{fig:PDCCoin} (bottom, left), we applied this approach, $\min\{\mathrm{eig}(M)\}$, to assess quantum correlations between the output photons of a single PDC coin.

	It is important to emphasize that nonclassical correlations are not accessible via first-order terms, $\langle\hat n_i\rangle$, that are typically focused on in coherent and linear walks to probe first-order interference.
	By contrast, higher-order auto- and cross-correlations, e.g., $\langle \hat n_i^2\rangle$ and $\langle \hat n_i\hat n_{i'}\rangle$, form the basis to infer quantum interference at one and between multiple outputs.

\subsection{Entanglement verification}
\label{Subsec:Entanglement}

	Another archetypal quantum phenomena between multiple quantum particles or quantized modes is entanglement, which provides a central resource for the function of quantum communication protocols \cite{HHHH09}.
	To probe entanglement that is generated through a DGQW in arbitrary degrees of freedom, we can construct the following test operator (see Appendix \ref{App:Entanglement}) in terms of the total input photon number [see Eq. \eqref{eq:InverseTrafo}]:
	\begin{equation}
	\label{eq:TestOp}
	\begin{aligned}
		\hat L
		=\sum_{i}\hat a_i^\dag\hat a_i
		=&\sum_{j,j',i}\left(
			U_{j,i}U^\ast_{j',i}\hat b_j^\dag\hat b_{j'}
			+ V_{j,i}^\ast V_{j',i}\hat b_j\hat b_{j'}^\dag
		\right.
		\\
		&\left.
			-U_{j,i}V_{j',i}\hat b_j^\dag\hat b_{j'}^\dag
			-V^\ast_{j,i}U^\ast_{j',i}\hat b_j\hat b_{j'}
		\right),
	\end{aligned}
	\end{equation}
	where the latter expansion yields the quantum correlations between the output fields.
	For finding this output entanglement, we now seek the minimal expectation value of $\hat L$ for separable (i.e., nonentangled) output states;
	this separable bound is denoted as $g_{\min}$.
	Thus, satisfying the inequalities
	\begin{equation}
	\label{eq:EntanglementCond}
		\langle \hat L\rangle<g_{\min}
		\quad\Leftrightarrow\quad
		0<g_{\min}-\langle \hat L\rangle
	\end{equation}
	determines entanglement of the output states, which are locally described by $\hat b_j$ and $\hat b_j^\dag$ for the $j$th subsystem.

	For calculating the sought-after bound $g_{\min}$, the method of separability eigenvalue equations can be applied \cite{SV13}.
	For the class of operators under consideration, $\hat L$, those equations have been solved analytically, using field-quadrature operators and their conjugated momenta, and applied to experiments with highly multimode Gaussian quantum light \cite{GSVCRTF15,GSVCRTF16}.
	For our purpose, the solutions can be recast into the form
	\begin{equation}
		\label{eq:SepBound}
		g_{\min}=\frac{1}{2}\sum_{j}\left(
			\sqrt{
				(1+2[VV^{\ast\mathrm{T}}]_{j,j})^2
				-4|[UV^\mathrm{T}]_{j,j}|^2
			}
			-1
		\right)
	\end{equation}
	for our formulation in terms of creation and annihilation operators.
	It is noteworthy that this bound is attained for a squeezed state that is a tensor-product vector of all output subsystems, $\bigotimes_j|\psi_j\rangle$.
	See Appendix \ref{App:Entanglement} for technical details.

	In addition, it is also worth mentioning that a displacement $\langle \hat b_i\rangle\neq0$ does not affect the entanglement because it describes a local transformation \cite{SV13,GSVCRTF15}.
	Consequently, and without a loss of generality, we can substitute $\hat b_j$ by $\Delta\hat b_j=\hat b_j-\langle\hat b_j\rangle\hat 1$ in the above definition of $\hat L$, which especially preserves the separable bound $g_{\min}$ \cite{SV13}.
	Therefore, the expectation value of the displaced $\hat L$ is $\langle\hat L\rangle=\sum_i\langle (\Delta\hat a_i)^\dag(\Delta\hat a_i)\rangle=0$, using the input light from Eq. \eqref{eq:InputState} for which $\langle \hat a_i^\dag\hat a_i\rangle=|\alpha_i|^2$ and $\langle \hat a_i\rangle=\alpha_i=\langle \hat a_i^\dag\rangle^\ast$ hold true.
	Hence, our entanglement criterion simplifies in Eq. \eqref{eq:EntanglementCond} to $0<g_{\min}-0$.
	In fact, the initial choice of $\hat L$ was made in a manner to achieve this accessible entanglement metric for DGQWs;
	see also Refs. \cite{PF17,SW17} in this context.
	We also remark that the inequality cannot be satisfied using only BS-type operations since $g_{\min}$ in Eq. \eqref{eq:SepBound} is zero for $V=0$.
	We applied this approach in Fig. \ref{fig:PDCCoin} to assess the entanglement from a single PDC coin operation.

	In conclusion, we described general DGQW in terms of a general process matrix $T$, Eq. \eqref{eq:FundamentalInOut}, combining input creation and annihilation operators at the output.
	In particular, the submatrix $V$ that encompasses the nonlinear contribution was vital for accessing and assessing quantum properties, such as entanglement, squeezing, and particle-number correlations, in experiments.
	The common coherent walk is based on $V=0$, thus unable to dynamically produce any of those quantum features.
	Nevertheless, quantum effects in coherent walks can occur because of input nonclassicality.
	For example, it is well known that nonclassical photon-number inputs for a purely unitary evolution produce an equal amount of output entanglement \cite{VS14}.
	However, this output quantumness is a consequence of the distributed input nonclassicality and not a feature of the process itself \cite{SW20}.
	Here, we showed that, even for classical and vacuum inputs, DGQW processes (where $V\neq0$) can indeed dynamically produce quantum features.


\section{Dynamics of DGQWs}
\label{Sec:Evolution}

	After providing general and exact input-output relations, including experimentally accessible quantumness metrics for the full walk, we now proceed to describing the nonlinear quantum evolution in detail.
	For sake of comparing coherent walks with our DGQWs, we explore two specific models in which all coins are either BS-type or PDC-type operations, as motivated in Sec. \ref{Sec:Motivation}.
	We emphasize again that one aim is to understand the quantumness of a quantum simulator itself rather than following the common path of considering input states that already carry some quantum properties.
	Thus, the following analysis is paramount to fathom and apply our DGQW framework as a dynamic means for achieving and maintaining an advanced quantum performance.

	The model from the motivation of the classical random walk is considered with different coins that can take two values, $+$ and $-$.
	Thus, it is convenient to further separate the vector of bosonic ladder operators according to the coin,
	\begin{equation}
		\vec{\hat a}=\begin{bmatrix}
			\vec{\hat a}_+ \\ \vec{\hat a}_-
		\end{bmatrix},
		\quad\text{with}\quad
		\vec{\hat a}_\pm=[\hat a_{(x,\pm)}]_{x=0,\ldots,d-1},
	\end{equation}
	where $x$ indicates the different vertices on the closed, one-dimensional graph;
	see the top-left depiction in Fig. \ref{fig:motivation}.
	A separation into more than two vector-valued components can be done analogously for multilevel coins.

	As per usual, each iteration of a quantum walk consists of a coin and a step operation that depends on the coin value.
	These two operations are then repeated $t$ times, giving us the full evolution of the quantum fields that describe composite coin-position quantum system.
	In our scenario, a step one position forward, $x\mapsto x+1$, is given by the matrix
	\begin{equation}
	\label{eq:DefineSigma}
		\Sigma
		=(\delta_{x'=x+1\,\mathrm{mod}\,d})_{x',x=0,\ldots,d-1}
		=\begin{bmatrix}
			0 & \cdots & \cdots & 0 & 1
			\\
			1 & \ddots & & & 0
			\\
			0 & \ddots & \ddots & & \vdots
			\\
			\vdots & \ddots & \ddots & \ddots & \vdots
			\\
			0 & \cdots & 0 & 1 & 0
		\end{bmatrix},
	\end{equation}
	which also accounts for the periodic boundary condition, $(d-1)+1=0\,\mathrm{mod}\, d$.
	This applies to the coin modes with $c=+$.
	The backwards step for $c=-$ is consequently described by $\Sigma^\mathrm{T}$.
	Hence, the joined step operation reads
	\begin{equation}
	\label{eq:DefineSigmaPlusMinus}
		\begin{bmatrix}
			\vec{\hat a}_+ \\ \vec{\hat a}_- \\ \vec{\hat a}_+^\dag \\ \vec{\hat a}_-^\dag
		\end{bmatrix}
		\mapsto
		\underbrace{\begin{bmatrix}
			\Sigma & 0 & 0 & 0
			\\
			0 & \Sigma^\mathrm{T} & 0 & 0
			\\
			0 & 0 & \Sigma & 0
			\\
			0 & 0 & 0 & \Sigma^\mathrm{T}
		\end{bmatrix}}_{=\Sigma_{\pm1}=\mathrm{diag}(\Sigma,\Sigma^\mathrm{T},\Sigma,\Sigma^\mathrm{T})}
		\begin{bmatrix}
			\vec{\hat a}_+ \\ \vec{\hat a}_- \\ \vec{\hat a}_+^\dag \\ \vec{\hat a}_-^\dag
		\end{bmatrix}.
	\end{equation}
	This Gaussian operation is of a form where $U=\left[\begin{smallmatrix}\Sigma & 0\\0 & \Sigma^\mathrm{T}\end{smallmatrix}\right]$ and $V=0$ apply.
	Thus, this particular map can only distribute quantum features over neighboring positions in each step but cannot generate quantumness itself.

	By contrast, the coin operation can achieve such a quantumness generation if we chose a PDC-type coin, as demonstrated in our motivation and analyzed in Fig. \ref{fig:PDCCoin}.
	When considering the same PDC process as a coin for all positions, the $4d\times 4d$ matrix of those nonlinear coins reads
	\begin{equation}
	\label{eq:DefGammaPDC}
		\Gamma_\mathrm{PDC}=\begin{bmatrix}
			\mu\mathrm{id} & 0 & 0 & \nu\mathrm{id}
			\\
			0 & \mu\mathrm{id} & \nu\mathrm{id} & 0
			\\
			0 & \nu^\ast\mathrm{id} & \mu^\ast\mathrm{id} & 0
			\\
			\nu^\ast\mathrm{id} & 0 & 0 & \mu^\ast\mathrm{id}
		\end{bmatrix}.
	\end{equation}
	Since each block is diagonal, this coin mixes coin (i.e., polarization) modes at the same position.
	Clearly, $V=\nu\left[\begin{smallmatrix}0&\mathrm{id}\\\mathrm{id}&0\end{smallmatrix}\right]\neq0$ holds true for all nontrivial squeezing values, $0\neq|\nu|^2=\sinh^2\xi$ for $\xi\neq0$.
	By contrast, the linear, BS-type coin reads
	\begin{equation}
	\label{eq:DefGammaBS}
		\Gamma_\mathrm{BS}=\begin{bmatrix}
			\tau\mathrm{id} & \rho\mathrm{id} & 0 & 0
			\\
			-\rho^\ast\mathrm{id} & \tau^\ast\mathrm{id} & 0 & 0
			\\
			0 & 0 & \tau^\ast\mathrm{id} & \rho^\ast\mathrm{id}
			\\
			0 & 0 & -\rho\mathrm{id} & \tau\mathrm{id}
		\end{bmatrix},
	\end{equation}
	where $V=0$ applies.
	Considering our in-depth and broadly applicable analysis in Sec. \ref{Sec:GenModel}, we can already see at this stage that a quantum simulator that is based on the nonlinear coin is the operation that is capable of dynamically producing---and exploiting in applications---quantum effects.

	For both types of coins, we can now compute the $t$-fold iteration of firstly coin and secondly step operation to obtain the full propagator, $T=(\Sigma_{\pm}\Gamma)^t$.
	Because of the periodic structure, we utilize a Fourier ansatz in Appendix \ref{App:Propagators} to provide a closed formulas for $T$ for both coin configurations.
	Here, for the sake of exposition, we focus on real-valued coin parameters, i.e., $\mu=\mathrm{Re}\,\mu$, $\nu=\mathrm{Re}\,\nu$, $\tau=\mathrm{Re}\,\tau$, and $\rho=\mathrm{Re}\,\rho$, while positive and negative values might still be possible.
	Because of $\mu^2-\nu^2=1=\tau^2+\rho^2$, we can write
	\begin{equation}
		\mu=\cosh\xi,
		\nu=\sinh\xi,
		\tau=\cos\vartheta,
		\text{ and }
		\rho=\sin\vartheta,
	\end{equation}
	with a rotation angle $\vartheta$ and a squeezing parameter $\xi$ that define the splitting ratios, Eq. \eqref{eq:SplittingRatio}.
	In Appendix \ref{App:Propagators}, the solution for the general complex case can be found for completeness.

\subsection{Analytic solutions}
\label{Subsec:AnalyticSolve}

	We can now analyze the exact results for $T$, which describes the output quantum fields, after $t$ iterations for comparing BS and PDC coins.
	For this purpose, we firstly provide the results from Appendix \ref{App:Propagators} for the propagation before discussing the quantities the formulas include.
	In the nonlinear case, we get
	\begin{equation}
	\label{eq:DecompTPDC}
	\begin{aligned}
		T_\mathrm{PDC}
		=&\left(\Sigma_\pm\Gamma_\mathrm{PDC}\right)^t
		\\
		=&
		\mathrm{diag}(F^\ast,F,F,F^\ast)
		\\
		&\times
		\begin{bmatrix}
			A_\mathrm{PDC} & 0 & 0 & B_\mathrm{PDC}
			\\
			0 & A_\mathrm{PDC} & B_\mathrm{PDC} & 0
			\\
			0 & B_\mathrm{PDC}^\ast & A_\mathrm{PDC}^\ast & 0
			\\
			B_\mathrm{PDC}^\ast & 0 & 0 & A_\mathrm{PDC}^\ast
		\end{bmatrix}
		\\
		&\times
		\mathrm{diag}(F,F^\ast,F^\ast,F),
	\end{aligned}
	\end{equation}
	and, for the unitary coin, we find
	\begin{equation}
	\label{eq:DecompTBS}
	\begin{aligned}
		T_\mathrm{BS}
		=&\left(\Sigma_\pm\Gamma_\mathrm{BS}\right)^t
		\\
		=&
		\mathrm{diag}(F^\ast,F^\ast,F,F)
		\\
		&\times
		\begin{bmatrix}
			A_\mathrm{BS} & B_\mathrm{BS} & 0 & 0
			\\
			-B_\mathrm{BS}^\ast & A_\mathrm{BS}^\ast & 0 & 0
			\\
			0 & 0 & A_\mathrm{BS}^\ast & B_\mathrm{BS}^\ast
			\\
			0 & 0 & -B_\mathrm{BS} & A_\mathrm{BS}
		\end{bmatrix}
		\\
		&\times
		\mathrm{diag}(F,F,F^\ast,F^\ast).
	\end{aligned}
	\end{equation}
	Firstly, $F$ is the unitary $d\times d$ matrix of the discrete Fourier transform, mapping a position, $x$, to a wave number, $k$.
	Secondly, $A$ and $B$ (including the subscripts ``PDC'' and ``BS'') denote generally complex-valued and always diagonal $d\times d$ matrices.
	To further describe these $A$ and $B$ matrices, we thirdly introduce other diagonal matrices,
	\begin{equation}
	\label{eq:Xi}
	\begin{aligned}
		\Xi
		=&\mathrm{diag}\Big\{\ln\Big[
			\mu\cos(2\pi k/d)
		\\&\phantom{\mathrm{xxxxxxxx}}
			+\sqrt{\mu^2\cos^2(2\pi k/d)-1}
		\Big]\Big\}_{k=0,\ldots,d-1},
		\\
		\Theta
		=&\mathrm{diag}\Big\{\arg\Big[
			\tau\cos(2\pi k/d)
		\\&\phantom{\mathrm{xxxxxxxx}}
			+i\sqrt{1-\tau^2\cos^2(2\pi k/d)}
		\Big]\Big\}_{k=0,\ldots,d-1},
	\end{aligned}
	\end{equation}
	comprising the squeezing parameter and rotation angle for each wave number $k$ and using $\arg(e^{i\vartheta})=\vartheta$.
	We additionally define $\mathrm{Re}(\Omega)=\mathrm{diag}\{\cos(2\pi k/d)\}_{k=0,\ldots,d-1}$, $\mathrm{Im}(\Omega)=\mathrm{diag}\{\sin(2\pi k/d)\}_{k=0,\ldots,d-1}$, and $\Omega=\mathrm{Re}(\Omega)+i\mathrm{Im}(\Omega)$.
	Finally, we can express the matrices $A$ and $B$ for all times $t$ as follows:
	\begin{equation}
	\label{eq:ABmatrices}
	\begin{aligned}
		A_\mathrm{PDC}
		=&\cosh(t\Xi)+i\frac{\mu\sinh(t\Xi)}{\sqrt{\mu^2(\mathrm{Re}\,\Omega)^2-1}}\mathrm{Im}\,\Omega,
		\\
		B_\mathrm{PDC}
		=&\frac{\nu\sinh(t\Xi)}{\sqrt{\mu^2(\mathrm{Re}\,\Omega)^2-1}}\Omega,
		\\
		A_\mathrm{BS}
		=&\cos(t\Theta)+i\frac{\tau\sin(t\Theta)}{\sqrt{1-\tau^2(\mathrm{Re}\,\Omega)^2}}\mathrm{Im}\,\Omega,
		\\
		B_\mathrm{BS}
		=&\frac{\rho\sin(t\Theta)}{\sqrt{1-\tau^2(\mathrm{Re}\,\Omega)^2}}\Omega,
	\end{aligned}
	\end{equation}
	where all involved functions act entrywise on the $d$ diagonal matrix elements, and all off-diagonal elements are zero.
	(See Appendix \ref{App:Propagators} for technical details.)
	Now, $T_\mathrm{PDC}$ and $T_\mathrm{BS}$ describe the exact transformation of input fields to output fields according to Eq. \eqref{eq:FundamentalInOut} after any number $t$ of repetitions of coin and step operations.
	In the following, we discuss these analytic findings.

\subsection{Spectral properties and discussion}
\label{Subsec:SpectralSolve}

	From the overall decomposition of $T_\mathrm{BS}$ in Eq. \eqref{eq:DecompTBS}, we unsurprisingly observe that $V=0$ holds true for the unitary evolution $T_\mathrm{BS}$ for all times $t$.
	This shows that, at no time $t$, quantumness can be produced by this linear walk while general interferences are still possible for this coherent walk.
	By contrast, the form of $T_\mathrm{PDC}$ in Eq. \eqref{eq:DecompTPDC} shows that $V=\left[\begin{smallmatrix}0 & F^\ast B_\mathrm{PDC} F\\FB_\mathrm{PDC}F^\ast&0\end{smallmatrix}\right]\neq0$ generally applies to the nonlinear process.

\begin{figure}
	\includegraphics[width=\columnwidth]{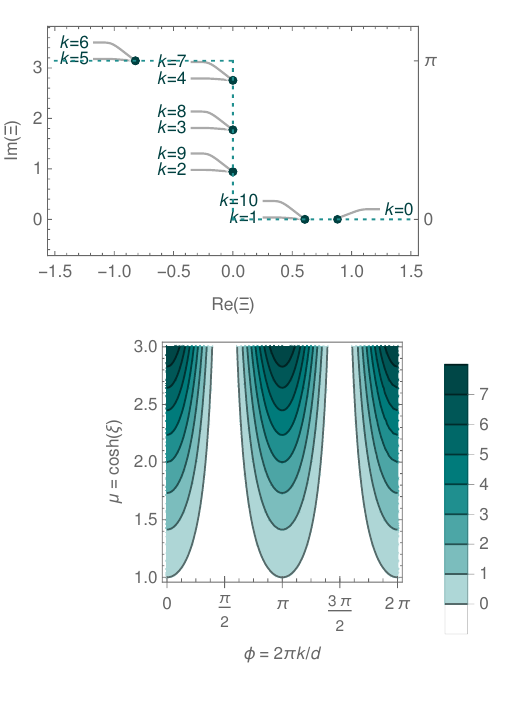}
	\caption{%
		Squeezing analysis for invariant wave-number-$k$ modes.
		The top plot shows the diagonal entries of $\Xi=\mathrm{diag}(\Xi_{0,0},\ldots,\Xi_{d-1,d-1})$ in Eq. \eqref{eq:Xi} for $\mu=\sqrt{2}$ (i.e., $\xi=\ln[\sqrt2+1]$ and $\nu=1$) and $d=11$.
		Regardless of the specific parameters, the real and imaginary parts are always found along the dashed line.
		A $\pi$-phase shift for $k$-number modes below the threshold $\mathrm{Re}(\Xi_{k,k})=0$ occurs.
		The bottom contour plot shows the general case as determined through the discriminant $\mu^2\cos^2\phi-1$.
		White areas describe $\Xi$'s diagonal elements that correspond to a pure rotation (no amplification for corresponding $k$ number, $\mathrm{Re}\,\Xi_{k,k}=0)$. 
	}\label{fig:spectal}
\end{figure}

	Furthermore, the rotation that is related to the BS-type coin can be expressed in the list of angles $\Theta$ for each wave number, Eq. \eqref{eq:Xi}, which act on the Fourier superposition modes that pick up a corresponding phase from each step operation.
	Similarly, the PDC coin results in various squeezing levels, collected in $\Xi$, that analogously apply to the Fourier modes.
	The squeezing parameters on the diagonal of $\Xi$ are depicted in the top plot of Fig. \ref{fig:spectal} for one example.
	Importantly, the rotation is governed by trigonometric functions, $\cos(t\Theta)$ and $\sin(t\Theta)$, while the squeezing operations are formulated in terms of hyperbolic rotations, $\cosh(t\Xi)$ and $\sinh(t\Xi)$; see Eq. \eqref{eq:ABmatrices}.

	The squeezing parameters in Eq. \eqref{eq:Xi} depend on but are not identical to the coin's squeezing parameter, $\mu=\cosh\xi=(e^{\xi}+e^{-\xi})/2$.
	For instance, the cosine of the phase $\phi=2\pi k/d$ is also relevant for determining the values in $\Xi$.
	Depending on $\phi$, the discriminant under the square root can be either positive (and zero) or negative---cf. bottom plot in Fig. \ref{fig:spectal}.
	In the former case, the output squeezing after $t$ iterations is, for example, described via $(e^{t\Xi_{k,k}}+e^{-t\Xi_{k,k}})/2$, in which each exponent is $t$-times scaled version of the squeezing value $\Xi_{k,k}$ that is obtained for the $k$th mode for $t=1$ combination of PDC coin and step.
	Thus, we have a $t$-fold increase of squeezing for that mode.
	When, however, the discriminant is negative, we have $\sqrt{\mu^2\cos^2\phi-1}=i\sqrt{1-\mu^2\cos^2\phi}$, a purely imaginary number.
	Then, $\ln(\mu\cos\phi+i\sqrt{1-\mu^2\cos^2\phi})=i\arg(\mu\cos\phi+i\sqrt{1-\mu^2\cos^2\phi})$ holds true, meaning that the mode for the correspond wave number $k$ is not squeezed but rotated instead, e.g., $\cosh(i\vartheta)=\cos(\vartheta)$.
	Also, a phase jump occurs (see Fig. \ref{fig:spectal}), resulting in a sign change, like $\cosh(\xi+i\pi)=-\cosh(\xi)$.
	Therefore, the nonlinear process includes signatures of rotations---being related to common coherence effects---and nonlinear quantum phenomena, e.g., squeezing.
	Depending on the input's field distribution, $j\mapsto\alpha_j$, we thus can generally produce a complex interplay of rotation and squeezing.
	In this context, it is also worth pointing out that position-localized inputs are in a superposition of all wave-number modes $k$.

	As a general remark beyond aforementioned specific evolution, we may comment on the general decomposition of a matrix $T$ in terms of $U$s and $V$s that obey Eq. \eqref{eq:UVconditions} and that describe the propagation at a time $t$ in arbitrary network geometries and possibly multivalued coins.
	Namely, the constraints in Eq. \eqref{eq:UVconditions} allows one to write
	\begin{equation}
		U=W\left[\mathrm{id}-SS^\ast\right]^{-1/2}
		\quad\text{and}\quad
		V=W\left[\mathrm{id}-SS^\ast\right]^{-1/2}S,
	\end{equation}
	where $Z^{-1/2}$ denotes a matrix for which $Z^{-1/2}ZZ^{-1/2}=\mathrm{id}$ holds true (using the usual matrix product), $S$ is a symmetric matrix ($S=S^\mathrm{T}$), and $W$ is a unitary rotation ($WW^{\ast\mathrm{T}}=\mathrm{id}$).
	Thus, if we were to interpret the results as scalars (e.g., via the diagonal matrix from Takagi's factorization of $S$), this would mean that $S$ yields a squeezing parameter, like $|S|=\tanh\zeta$, such that $\cosh\zeta=1/\sqrt{1-|S|^2}$ and $\sinh\zeta=|S|/\sqrt{1-|S|^2}$ describe Lorentz boosts in Minkowski space.
	Therefore, any generic DGQW may include both rotations, via $W$, and hyperbolic rotations, through $S$.
	See also Refs. \cite{S18,ACLS13} for related transformation properties in quantum walks.

	In summary, we analyzed the temporal properties of the nonlinear evolution in this section.
	Together with the results from Sec. \ref{Sec:GenModel}, this allows us to characterize the general quantum-dynamical behavior of DGQWs.
	In the following, we can now jointly apply all our general findings to investigate specific examples of the nonlinear quantum evolution of DGQWs for all times.

\begin{figure*}
	\includegraphics[width=\textwidth]{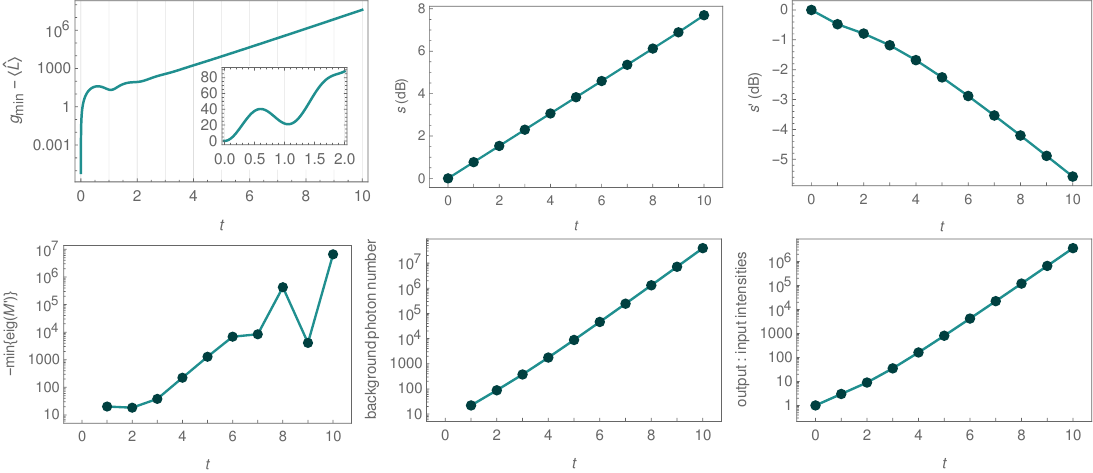}
	\caption{%
		Quantum and nonlinear characteristics of a DGQW on a $d=11$-periodic position space (cf. top-left graph in Fig. \ref{fig:motivation}) with a polarization PDC coin at each vertex that is characterized by $\mu=\sqrt{2}$ and $\nu=1$---likewise, a coin squeezing parameter $\xi=\ln(\sqrt{2}+1)$ and a biased splitting ratio $R_\mathrm{PDC}=1/2$.
		The classical coherent input state is given through localized coherent amplitudes, $\alpha_{(0,+)}=\sqrt{10}$, $\alpha_{(0,-)}=i\sqrt{10}$, and zero otherwise.
		The first plot (top, left) shows the generated entanglement.
		To exemplify that our solutions apply to intermediate times as well, the full interval $0\leq t\leq 10$ of the continuous evolution is depicted.
		The inset highlights oscillations for small times.
		The center-top plot shows the multimode squeezing of the discrete-time DGQW.
		Interestingly, when restricting to the squeezing between polarizations only, top-right plot, we observe only antisqueezing that is not a quantum signature and shows fluctuations well above the vacuum limit.
		The bottom-left plot certifies the nonclassical photon-number correlations between the quantum particle, regardless of their polarization.
		The background noise, summed over all positions and coin values, is shown in the center-bottom plot.
		Lastly, the overall amplification of the nonlinear process, while omitting background noise, relative to the total input photon number is provided (bottom, right).
	}\label{fig:application}
\end{figure*}


\section{Application}

	We now employ the collection of all our  general findings from Secs. \ref{Sec:GenModel} and \ref{Sec:Evolution} to study specific DGQWs in greater detail.
	In Fig. \ref{fig:application}, an overview for one scenario is provided.
	Therein, we consider the quantum and nonlinear properties of the DGQW evolution for classical input light, which we discuss in detail in the continuation of this section.

\subsection{Quantum features}

	Firstly, we can consider the generation of entanglement via DGQWs from a fully factorizable input state, Eq. \eqref{eq:InputState}, initially located at position $x=0$.
	The top-left plot in Fig. \ref{fig:application} shows the successive---in fact, exponential---buildup of highly multimode entanglement, using the entanglement criterion from Eq. \eqref{eq:EntanglementCond}.
	While a single PDC coin only produces entanglement between the polatization (likewise, coin) modes at each point $x$ (see Sec. \ref{Sec:Motivation}), the step operation then distributes this entanglement to neighboring positions, $x\pm1$.
	With each application of coin and step operation, $t\to t+1$, the entanglement is then reestablished between the polarizations at each position and farther distributed across all positions.

	Please note that the depiction of the entanglement evolution in Fig. \ref{fig:application} is done for a continuous time $t$ to emphasize that approach applies to continuous evolution and discrete-time walks alike.
	For this purpose, the $4d\times 4d$ generator $G$ of the evolution is given by the matrix logarithm that is obtainable through the spectral decomposition considerations from Appendix \ref{App:Propagators}.
	This results in Heisenberg equations of motion,
	\begin{equation}
		\partial_t\begin{bmatrix}
			\vec{\hat b}(t)
			\\
			\vec{\hat b}(t)^\dag
		\end{bmatrix}
		=G\begin{bmatrix}
			\vec{\hat b}(t)
			\\
			\vec{\hat b}(t)^\dag
		\end{bmatrix}
	\end{equation}
	for bosonic field operators, with $G=\ln(\Sigma_\pm\Gamma_\mathrm{PDC})$ and the initial value $\vec{\hat b}(0)=\vec{\hat a}$, being solved by $T_\mathrm{PDC}$ from Sec. \ref{Subsec:AnalyticSolve}.

	Secondly, and coming back to our discrete-time walk, squeezing is studied as a quantum phenomena of light that can manifest itself in a single and multiple modes.
	The top-center graph in Fig. \ref{fig:application} shows the multimode squeezing that is obtained from the full, propagated covariance matrix $C$, as described in Sec. \ref{Sec:GenModel}.
	In addition, we can analyze the squeezing when restricting ourselves to a single position, resulting in the reduced $4\times 4$ covariance matrix
	\begin{equation}
	\begin{aligned}
		&C'
		\\=&\begin{bmatrix}
			\frac{1}{2}\langle \{(\Delta \hat q_{x,c})(\Delta \hat q_{x,c'})\}\rangle
			&
			\frac{1}{2}\langle \{(\Delta \hat q_{x,+})(\Delta \hat p_{x,c'})\}\rangle
			\\
			\frac{1}{2}\langle \{(\Delta \hat p_{x,c})(\Delta \hat q_{x,c'})\}\rangle
			&
			\frac{1}{2}\langle \{(\Delta \hat p_{x,+})(\Delta \hat p_{x,c'})\}\rangle
		\end{bmatrix}_{c,c'\in\{+,-\}}
	\end{aligned}
	\end{equation}
	for the coin.
	Note that this matrix is identical for all positions $x$ since the coin is the same for all vertices of the considered graph.
	The squeezing between the $+$ and $-$ polarization at each position is shown in the top-right plot in Fig. \ref{fig:application}.
	In contrast to our considerations in Sec. \ref{Sec:Motivation} where the single coin did lead to squeezing, we here observe no squeezing (negative values) in the polarization.
	This is because each coin application is followed by a step operation, which distributes this coin's squeezing to neighboring positions, as discussed for entanglement.
	Therefore, the top-center plot in Fig. \ref{fig:application} verifies that we have a nonlocalized squeezing.

	Thirdly, we characterize photon-number correlations in Fig. \ref{fig:application} (bottom, left).
	Since a polarization-resolved measurement at each output position might not be possible in all experiments, we here consider the joint photon number of both modes, which can be formulated through a transformation with a rectangular $d\times 2d$ matrix, e.g.,
	\begin{equation}
		\begin{bmatrix}
			\mathrm{id} & \mathrm{id}
		\end{bmatrix}
		\begin{bmatrix}
			\langle \vec{\hat n}_+\rangle
			\\
			\langle \vec{\hat n}_-\rangle
		\end{bmatrix}
		=\langle \vec{\hat n}_+\rangle+\langle \vec{\hat n}_-\rangle.
	\end{equation}
	Similarly, the matrix of normally ordered moments can be transformed in this manner, $M'=\left[\begin{smallmatrix}\mathrm{id}&\mathrm{id}\end{smallmatrix}\right]M\left[\begin{smallmatrix}\mathrm{id}&\mathrm{id}\end{smallmatrix}\right]^\mathrm{T}$, to quantify particle-number correlations independently of the coin degree of freedom.
	The result of this treatment in shown in the bottom-left plot of Fig. \ref{fig:application}.
	Note that, in contrast to Fig. \ref{fig:PDCCoin}, the negative negativity is depicted in Fig. \ref{fig:application}, meaning that a positive value implies nonclassical correlation.
	Here, nonclassical correlations in positions, being the remaining degree of freedom in $M'$, are verified.
	Compared to the other quantumness metrics, however, this nonclassicality is not simply monotonously increasing;
	rather, it oscillates, being best visible for later times $t$.

\subsection{Nonlinear effects}

	As discussed previously, it makes also sense to study the nonlinear effects that originate from the PDC coin.
	Beyond that, we want to analyze the quantum-nonlinear properties of the evolution.
	Firstly, this enables us tell the different effects apart that are described by nonlinear PDC-based maps and are not expected from linear BS-like transformations (cf. Sec. \ref{Sec:Evolution}).
	Secondly, the difference between classical coherence in nonlinear optics and nonlinear quantum effects can be farther analyzed in this manner, complementing the quantification of quantum effects in the previous subsection.
	To this end, equations of motions for the classical and nonclassical dynamics can be formulated for general nonlinear interactions \cite{SW20} whose solutions unveil the distinct differences between a classical and quantum evolution.
	Thereby, the quantumness of the process is investigated.

	Using the results from Ref. \cite{SW20}, we find that the classical evolution of a second-order nonlinearity, such as PDC in the quantum domain, can be fully described by $\vec \beta$ as a function of $T$ (thus, a function of $t$) and the input $\vec \alpha$, Eq. \eqref{eq:AmplitudePropagation}, which is equivalent to ignoring all quantum-physical commutation relations.
	Since the background noise stems from commutators (Sec. \ref{Sec:Motivation}), the overall background noise can be computed as
	\begin{equation}
		\sum_{x,c}\left(
			\langle \hat n_{x,c}\rangle-|\beta_{x,c}|^2
		\right).
	\end{equation}
	Thereby, we describes this quantum-physical consequence of the nonlinear process under study when compared to the classical, yet still nonlinear evolution.
	The successive increase of the overall background noise is shown in Fig. \ref{fig:application} (center, bottom).
	Please also note that the background noise---again, because of homogeneity of the coins at all vertices of the graph that defines the positions---is independent of the position $x$ and already present in the DGQW depicted in Fig. \ref{fig:motivation}.

	Another consequence of the nonlinear PDC process we discussed in Sec. \ref{Sec:Motivation} is amplification.
	To assess the gain in the signal alone, being applicable to the quantum dynamics and the classical nonlinear evolution, we can consider the total photon number without the excess noise, being normalized to the total input photon number, $\sum_{x,c}|\beta_{x,c}|^2/\sum_{x,c}|\alpha_{x,c}|^2$.
	This is shown in the bottom-right plot in Fig. \ref{fig:application}.
	Clearly, with each iteration $t\to t+1$, the signal is exceedingly amplified with the help of the nonlinear PDC coin.

\subsection{Combined coins}

	As a final example, we may combine coins based on PDC and BS operations to show the general applicability of the methodology introduced in this work.
	This coin combination results in a DGQW that produces the first-order interferences in the position-based and time-dependent photon-number distribution, as seen for classical coherent walk in Fig. \ref{fig:motivation}, together with the quantum characteristics of PDC-based walks we just discussed.

\begin{figure}
	\includegraphics[width=\columnwidth]{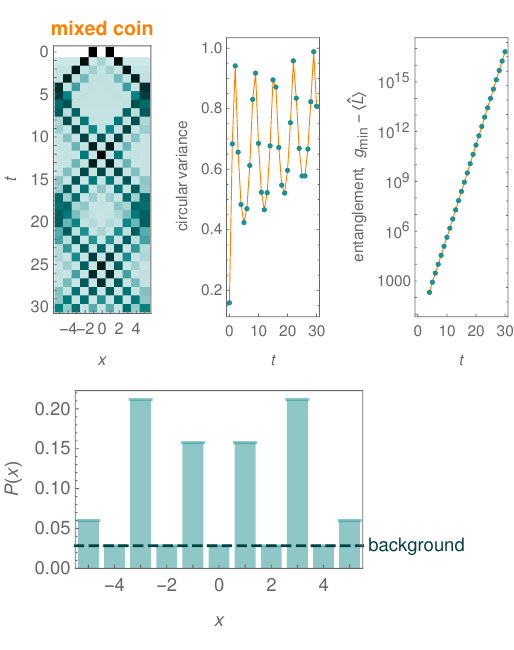}
	\caption{%
		DGQW in which a PDC coin, with $\mu=\sqrt{2}$ and $\nu=1$, is mixed with a BS coin, $\tau=1/\sqrt2=\rho$, followed by the $d$-periodic position change, $\Sigma_{\pm1}$.
		This scenario combines the characteristic interference patterns of a coherent walk (cf. Fig. \ref{fig:motivation}), e.g., quantified through the circular variance, with quantum superposition effects, e.g., the depicted entanglement, of a genuine quantum walk (cf. Fig. \ref{fig:application}).
		What is also different to our previous examples is that we here chose inputs distributed over different positions, $\alpha_{(x=1,c=+)}=\sqrt{10}$ and $\alpha_{(x=-1,c=-)}=i\sqrt{10}$ and $\alpha_{(x,c)}=0$ otherwise.
		The bottom plot exemplifies one normalized intensity distribution---labeled as $P(x)$ and being distinctively different from an incoherent random walk---for the $15$th iteration (i.e., $t=15$) of coins and step operations.
		The excess noise of the nonlinear quantum propagation is indicated as a dashed horizontal line.
	}\label{fig:mixedcoins}
\end{figure}

	In Fig. \ref{fig:mixedcoins}, we apply a PDC coin first, before applying a BS-type coin, jointly described through the coin map
	\begin{equation}
		\Gamma_\mathrm{BS}\Gamma_\mathrm{PDC}
		=\begin{bmatrix}
			\mu\tau\mathrm{id} & \mu\rho\mathrm{id} & \nu\rho\mathrm{id} & \nu\tau\mathrm{id}
			\\
			-\mu\rho^\ast\mathrm{id} & \mu\tau^\ast\mathrm{id} & \nu\tau^\ast\mathrm{id} & -\nu\rho^\ast\mathrm{id}
			\\
			\nu^\ast\rho^\ast\mathrm{id} & \nu^\ast\tau^\ast\mathrm{id} & \mu^\ast\tau^\ast\mathrm{id} & \mu^\ast\rho^\ast\mathrm{id}
			\\
			\nu^\ast\tau\mathrm{id} & -\nu^\ast\rho\mathrm{id} & -\mu^\ast\rho\mathrm{id} & \mu^\ast\tau\mathrm{id}
		\end{bmatrix}.
	\end{equation}
	In contrast to the PDC coin alone, cf. Eq. \eqref{eq:DefGammaPDC}, this coin also mixes $\hat a_{x,+}$ and $\hat a_{x,-}$.
	And, contrasting the BS coin in Eq. \eqref{eq:DefGammaBS}, it too superimposes $\hat a_{x,+}$ and $\hat a_{x,-}^\dag$.
	For the specific example in Fig. \ref{fig:mixedcoins}, we choose splitting ratios $R_\mathrm{PDC}=1/2$ (i.e., $1:2$) and $R_\mathrm{BS}=1$ (a fair coin) between transmitted and reflected light, bringing together the Hadamard coherent walk in Fig. \ref{fig:motivation} and the biased-coin DGQW in Fig. \ref{fig:application}.

	With this combination of coins, all kinds of classical and quantum interference effects discussed throughout this work are jointly governing the DGQW in Fig. \ref{fig:mixedcoins}.
	(Note that, compared to previous examples, the step operation $\Sigma_\pm$ was not altered for consistency, but the input state was changed for variety.)
	Particularly, we show the intensity oscillation in the $d$-periodic graph, quantified through the circular variance, and the ever increasing amount of entanglement.
	Thereby, we combine classical and quantum coherence phenomena in our framework.
	More generally, a general DGQW, mainly defined by $V\neq0$, thus renders a variety of quantum simulations possible, ranging from classical wave phenomena over propagating quantum particles to the continuous evolution of quantized fields.

\begin{figure*}
	\includegraphics[width=\textwidth]{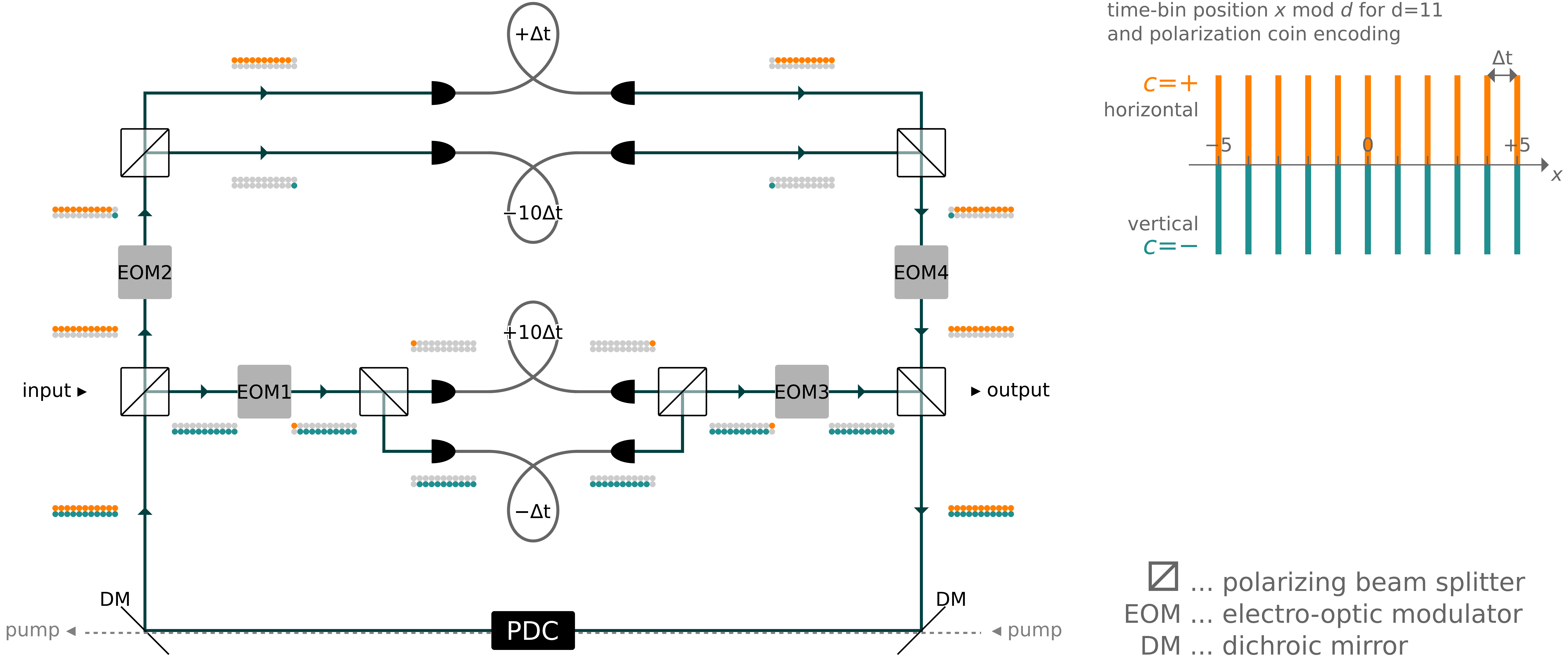}
	\caption{%
		Proposed scheme for a DGQW on a closed graph with $d=11$ vertices, as depicted in Fig. \ref{fig:motivation}.
		The setup on the left describes a looped Mach--Zehnder interferometer \cite{SCPGMAJS10} with a single PDC crystal for all coin operations.
		So-called time bins encode the position $x$ in temporally displaced light pulses (right-top plot).
		Differently shaded circles in the left plot indicate the action of electro-optic modulators (EOMs), polarizing BSs, and fibers.
		The latter introduce relative delays in terms of multiples of $\Delta t$ that separate pulses in time bins.
		The bottom PDC process is driven by a pulsed laser that acts on each time bin and which is seeded via the incident polarization (i.e., coin) states.
	}\label{fig:expscheme}
\end{figure*}

\subsection{Beyond Gaussian walks}

	In this work, we focused on nonlinear Gaussian processes.
	Here, let us briefly outline how out methodology can be extended to higher-order nonlinearities.
	Non-Gaussian elements are, for instance, important for implementing universal continuous-variable quantum information processing \cite{LB99}, including the possibility to employ non-Gaussian input states \cite{KLM01} and non-Gaussian detection schemes \cite{BS02}.

	For a universal quantum simulator, which is one of the main applications of quantum walks, the same requirement for at least one non-Gaussian component applies.
	Here, for instance, we considered photon-number measurements for DGQWs, including their nonclassical correlations (i.e., fourth-order field correlations), to satisfy the non-Gaussian demand \cite{BS02} beyond balanced homodyne detection which solely addresses the covariance matrix of quadratures and their momenta.
	From our methodology, we can also straightforwardly conclude non-Gaussian inputs, such as photon-number states \cite{KLM01}, as a non-Gaussian option.
	Specifically, an $n$-photon Fock state can be expressed via coherent states $|\alpha\rangle$ via the relation $(n!)^{-1}(\partial_{\alpha}\partial_{\alpha^\ast})^n[e^{|\alpha|^2}|\alpha\rangle\langle\alpha|]_{\alpha=0}$ (see, e.g., Ref. \cite{VS14}), which analogously applies to multimode number states.
	Hence, we can apply the same derivatives to expectation values of output modes for coherent input states we already computed to predict the behavior of number states at the input of DGQWs.

	Besides non-Gaussian measurements and input states, one can also include other nonlinear processes in the walk.
	Like with our replacement of BS coin maps to arrive at DGQWs (cf. Sec. \ref{Sec:Motivation}), one can imagine cross-Kerr effect as a higher-order nonlinearity as an example, which is modeled by an fourth-order interaction Hamilton that is proportional to $\hat n_+\hat n_{-}=\hat b_+^\dag\hat b_-^\dag\hat b_+\hat b_-$.
	See Ref. \cite{MDB15} for a related scheme.
	This results in the highly nonlinear input-output relation $\hat b_{\pm}=\exp[i\varphi\hat n_{\mp}]\hat a_{\pm}$ for the two coin components, correlating a field with a phase that depends on the intensity of the other field.
	In Ref. \cite{NPR07}, it was proposed to replace the linear medium in which the photons propagate during the walk with a nonlinear (single-mode) Kerr medium, with a Hamiltonian $\propto \hat n_\pm^2$.
	By contrast, the cross-nonlinearity here is essential for the function of the quantum walk itself, rather than constituting a nonlinear background medium.

\subsection{Proposed experimental scheme for DGQWs}

	To conclude our work that introduces the concept of DGQWs, we propose an experimental implementation.
	The specific setup we put forward is a combination of a well-established time-bin-based quantum walk architecture (see, e.g., Ref. \cite{SCPGMAJS10}) and cutting-edge loop-operated PDC experiments \cite{ESTBDPBS21}.
	This time-multiplexing scheme is depicted in Fig. \ref{fig:expscheme}.
	Therein, an interferometer (top arms) is looped back onto itself via the bottom feedback line that includes the PDC coin.

	We begin with the step operation.
	Positions $x$ are defined in terms of time bins, containing pulses of light as the walker, and the coin state is encoded in the polarization, as assumed throughout this article.
	Depending on the coin state $c=\pm$, the light pulse of the $x$th bin propagates through either the top two or middle two arms of the interferometer.
	Because of the length difference of these arms (Fig. \ref{fig:expscheme}), a relative temporal shift of $\pm\Delta t$ is picked up for both polarizations, resulting in a $x\pm1$ position change and realizing the step operation \cite{SCPGMAJS10}.
	To connect positions $x=0$ and $x=d-1$, two fibers with a larger length differentials, $\pm (d-1)\Delta t$, are used and actively routed to other fibers by EOMs.
	That is, the EOM is triggered such that pulses for positions $x=0$ and $x=d-1$ experience an position offset of $x\pm (d-1)\,\mathrm{mod}\,d$.
	This renders it possible to achieve the periodic boundary conditions for the graph depicted in the top-left plot of Fig. \ref{fig:motivation}.

	The bottom arm in Fig. \ref{fig:expscheme}, also connecting output and input, yields an actual time step.
	Within this arm, we also find the PDC coin operation, replacing the common waveplate-based BS coin of our earlier experiments.
	The PDC process can be driven by pulse trains from a pump laser, leading to the seeded (i.e., stimulated) emission of signal and idler photons in the two polarizations that are characteristic for second-order nonlinear processes.
	(Note that the pump light does not further contribute to the quantum walk, as ensured by dichroic mirrors in Fig. \ref{fig:expscheme}.)
	We recently demonstrated that it is feasible to reliably and coherently operate PDC process in a loop-based configuration in such a manner \cite{ESTBDPBS21}.
	In addition, it is noteworthy that deterministic in- and out-couplings can be realized in such otherwise self-contained (i.e., looped) interferometer by virtue of the EOMs in Fig. \ref{fig:expscheme};
	see Ref. \cite{NBKSSGPKJS18} for details.

	With the proposed setup, DGQW may be explored in future experiments.
	As we laid out, the individual components of the scheme are available, such as time-multiplexed quantum walks and operating a PDC element in loops, in our own experiments.


\section{Conclusion}

	In summary, we introduced and explored the concept of DGQWs.
	We in particular showed that this type of quantum simulator does produce genuine quantum effects over its nonlinear evolution, setting it apart from linear quantum walks and even coherent walks.

	In the first step, we formulated the notion of a coherent walk, an intermediate stage between classical random walks and genuine quantum walks.
	Specifically, superposition---regardless of the classical or quantum origin---suffice for a coherent walk, which are accessible, for instance, with classical optics.
	Thus, a coherent walk goes beyond a random walk but not necessarily requires quantum superpositions.
	By contrast, a genuine quantum walk requires a quantum-physical treatment.
	This also means that the evolution itself is the origin of quantum effects, and not the input state's features.
	We proved that a DGQW satisfies these demands.
	In addition, a DGQW is an example of a nonlinear walk, contrasting the most frequently considered scenario of linear walks.
	For example, a DGQW actively produces quantum particles in its evolution because of stimulated emission of photon pairs whose quantum properties are even further amplified with each subsequent iteration of steps.
	Thus, even vacuum can constitute a useful input state that produces pronounced quantum effects.

	Prominent examples of quantum phenomena we investigated are squeezing, particle-number correlations, and entanglement.
	Those effects are highly relevant in experiments and applications, e.g., squeezing for quantum metrology and entanglement for quantum communication.
	We provided measurable criteria to verify the presence---as well as the buildup---of such quantum phenomena over the DGQW's evolution.
	Moreover, the nonlinear nature of the process under study offered access to nonlinear quantum-optical effects, too, such as amplification that results in comparably bright output light.
	We also characterized the excess noise as an imperfection that is a joint result of the nonlinear amplification and vacuum quantum fluctuations.

	We compared DGQWs with prime examples of coherent walks, especially quantum-optical networks with coherent light and linear optical components as coin operators.
	Thereby, we proved the superior quantum performance of DGQWs, offering a quantum evolution that is not accessible with classical interference alone.
	We also found that a coherent walk can be considered as a special instance of our general DGQW framework.
	For the sake of exposition, our focus was on replacing the coin operation from a coherent walk with a nonlinear coin.
	But the broader framework introduced here even allows for considering nonlinear processes at any point of the quantum walk, including the possibility of a nonlinear step operation.
	Moreover, the general methodology is not limited to processes that are homogeneous in time, simple one-dimensional geometries, two-level coins, etc.
	For example, changing the driving pump of the process that governs the nonlinear process renders it possible to actively control the quantum light's evolution.
	By ignoring quantum-physical commutation relations, we were additionally able to compare nonlinear, but only coherent walks with our DGQW.

	The model devised in this work is mainly based on the Gaussian formalism for bosonic multimode quantum fields.
	Although we concentrated on relatively pure and ideal scenarios for the purpose of the introduction of the concept presented here, the formalism allows one to describe experimentally relevant impurities (e.g., losses, thermal noise, etc.) rather straightforwardly \cite{WPGCRSL12,ARL14}.
	In fact, we previously considered how losses and nonlinear processes interact \cite{ELDBDPSS20}.
	For instance, one can use that approach to balance loss and quantum amplification, while keeping excess noise low.
	Additionally, a comprehensive study of imperfections in related Gaussian boson sampling scenarios can be found in Ref. \cite{PWRWTS19}.

	The continuous-variable approach that is at the heart of our DGQW methodology also renders few-photon approximations superfluous.
	The involved Hilbert state is of infinite dimensionality and can consist of an arbitrary number of quantized modes.
	By adapting our time-multiplexed quantum walk platform, we proposed a scheme for implementing the specific scenario under study, harnessing a PDC coin in a looped Mach--Zehnder interferometer.
	We also foresee that a promising architecture for implementing DGQWs are nonlinear waveguide arrays \cite{Setal14,KKCSWLGHJS13} that can embed the linear and nonlinear elements rather efficiently.
	This renders our approach practicable with state-of-the-art quantum devices.

	Therefore, a versatile framework for nonlinear quantum walks has been established and explored in this work.
	This allows one to tap this potential to study nonlinear quantum-physical transport phenomena.
	Also, dynamical quantum communication scenarios in complex networks can benefit from the nonlinear character of the evolution and the continuous buildup of quantum correlations reported here.


\begin{acknowledgments}
	The authors thank Matthias Pukrop for valuable comments.
	The Integrated Quantum Optics group acknowledges financial support through the European Commission through the ERC project QuPoPCoRN (Grant No. 725366).
	J. S. acknowledges financial support from the Deutsche Forschungsgemeinschaft (DFG, German Research Foundation) through the Collaborative Research Center TRR 142 (Project No. 231447078, project C10).
\end{acknowledgments}

\appendix

\section{Field moments and their transformations}
\label{App:MomentTransform}

	In this appendix, we formulate the exact input-output relations for different properties of quantum field, especially their moments.
	For convenience, we introduce the abbreviations
	\begin{equation}
		\hat u_i=\sum_j U_{i,j} \hat a_j
		\quad\text{and}\quad
		\hat w_i=\sum_j V_{i,j}^\ast \hat a_j
	\end{equation}
	such that we can decompose the output annihilation operator as $\hat b_i=\hat u_i+\hat w_i^\dag$.
	The following commutation relations apply to this representation:
	\begin{equation}
	\begin{aligned}
		\hat u_{i}\hat u_{i'}^\dag-\hat u_{i'}^\dag\hat u_{i}
		=&x_{i,i'}=\sum_{j}U_{i,j}U_{i',j}^\ast=x_{i',i}^\ast,
		\\
		\hat w_{i}\hat w_{i'}^\dag-\hat w_{i'}^\dag\hat w_{i}
		=&y_{i,i'}=\sum_{j}V_{i,j}^\ast V_{i',j}=y_{i',i}^\ast,
		\\
		\hat u_{i}\hat w_{i'}^\dag-\hat w_{i'}^\dag\hat u_{i}
		=&z_{i,i'}=\sum_{j}U_{i,j}V_{i',j}=z_{i',i}.
		\\
		\hat w_{i}\hat u_{i'}^\dag-\hat u_{i'}^\dag\hat w_{i}
		=&z_{i',i}^\ast=\sum_{j}V_{i,j}^\ast U_{i',j}^\ast=z_{i,i'}^\ast,
	\end{aligned}
	\end{equation}
	also introducing the symbols $x$, $y$, and $z$ which essentially are the entries of the products of the transformation matrices $U$ and $V$, thus too obeying their properties in Eq. \eqref{eq:UVconditions}.
	Applying the above commutation relations, we obtain second-order output field correlations,
	\begin{equation}
	\begin{aligned}
		\hat b_{i}\hat b_{i'}
		=&\hat u_{i}\hat u_{i'}+\hat w_{i}^\dag\hat u_{i'}+\hat w_{i'}^\dag\hat u_{i}+\hat w_{i}^\dag\hat w_{i'}^\dag+z_{i,i'},
		\\
		\hat b_{i}^\dag\hat b_{i'}
		=&\hat u_{i}^\dag\hat u_{i'}+\hat w_{i}\hat u_{i'}+\hat u_{i}^\dag\hat w_{i'}^\dag+\hat w_{i'}^\dag\hat w_{i}+y_{i,i'}.
	\end{aligned}
	\end{equation}
	In particular, the equality for $i=i'$ describes the output photon-number operator for the $i$th mode, $\hat b_{i}^\dag\hat b_{i}=\hat u_{i}^\dag\hat u_{i}+\hat w_{i}\hat u_{i}+\hat u_{i}^\dag\hat w_{i}^\dag+\hat w_{i}^\dag\hat w_{i}+y_{i,i}$.
	The normally ordered second-order photon-number correlation is given by the fourth-order term
	\begin{equation}
	\begin{aligned}
		&\hat b_{i}^\dag\hat b_{i'}^\dag\hat b_{i}\hat b_{i'}
		\\
		=&|z_{i,i'}|^2+y_{i,i}y_{i',i'}+y_{i,i'}y_{i',i}
		\\
		&+z_{i,i'}^\ast(\hat u_{i}\hat u_{i'}+\hat w_{i}^\dag\hat u_{i'}+\hat w_{i'}^\dag\hat u_{i}+\hat w_{i}^\dag\hat w_{i'}^\dag)
		\\
		&+z_{i,i'}(\hat u_{i}^\dag\hat u_{i'}^\dag+\hat u_{i'}^\dag\hat w_{i}+\hat u_{i}^\dag\hat w_{i'}+\hat w_{i}\hat w_{i'})
		\\
		&+y_{i,i}(\hat u_{i'}^\dag\hat u_{i'}+\hat u_{i'}^\dag\hat w_{i'}^\dag+\hat w_{i'}\hat u_{i'}+\hat w_{i'}^\dag\hat w_{i'})
		\\
		&+y_{i',i'}(\hat u_{i}^\dag\hat u_{i}+\hat u_{i}^\dag\hat w_{i}^\dag+\hat w_{i}\hat u_{i}+\hat w_{i}^\dag\hat w_{i})
		\\
		&+y_{i,i'}(\hat u_{i'}^\dag\hat u_{i}+\hat u_{i'}^\dag\hat w_{i}^\dag+\hat w_{i'}\hat u_{i}+\hat w_{i'}^\dag\hat w_{i})
		\\
		&+y_{i',i}(\hat u_{i}^\dag\hat u_{i'}+\hat u_{i}^\dag\hat w_{i'}^\dag+\hat w_{i}\hat u_{i'}+\hat w_{i}^\dag\hat w_{i'})
		\\
		&+\hat u_{i}^\dag\hat u_{i'}^\dag\hat u_{i}\hat u_{i'}
		+\hat u_{i}^\dag\hat u_{i'}^\dag\hat w_{i}^\dag\hat u_{i'}
		+\hat u_{i}^\dag\hat u_{i'}^\dag\hat w_{i'}^\dag\hat u_{i}
		+\hat u_{i}^\dag\hat u_{i'}^\dag\hat w_{i}^\dag\hat w_{i'}^\dag
		\\
		&+\hat u_{i'}^\dag\hat w_{i}\hat u_{i}\hat u_{i'}
		+\hat u_{i'}^\dag\hat w_{i}^\dag\hat w_{i}\hat u_{i'}
		+\hat u_{i'}^\dag\hat w_{i'}^\dag\hat w_{i}\hat u_{i}
		+\hat u_{i'}^\dag\hat w_{i}^\dag\hat w_{i'}^\dag\hat w_{i}
		\\
		&+\hat u_{i}^\dag\hat w_{i'}\hat u_{i}\hat u_{i'}
		+\hat u_{i}^\dag\hat w_{i}^\dag\hat w_{i'}\hat u_{i'}
		+\hat u_{i}^\dag\hat w_{i'}^\dag\hat w_{i'}\hat u_{i}
		+\hat u_{i}^\dag\hat w_{i}^\dag\hat w_{i'}^\dag\hat w_{i'}
		\\
		&+\hat w_{i}\hat w_{i'}\hat u_{i}\hat u_{i'}
		+\hat w_{i}^\dag\hat w_{i}\hat w_{i'}\hat u_{i'}
		+\hat w_{i'}^\dag\hat w_{i}\hat w_{i'}\hat u_{i}
		+\hat w_{i}^\dag\hat w_{i'}^\dag\hat w_{i}\hat w_{i'}.
	\end{aligned}
	\end{equation}
	Please also note that quadrature- and momentum-based expressions are obtained from the above calculations as well because of $\hat q_i=(\hat b_i+\hat b_i^\dag)/\sqrt{2}$ and $\hat p_i=(\hat b_i-\hat b_i^\dag)/(i\sqrt{2})$.

	For a coherent input state, which is studied throughout this work, $\bigotimes_j|\alpha_j\rangle$, the above expressions reduce to the following output expectation values:
	\begin{equation}
	\begin{aligned}
		\langle \hat b_{i} \rangle
		=&\langle \hat u_{i}\rangle+\langle \hat w_{i}^\dag\rangle =\sum_{j}(U_{i,j}\alpha_j+V_{i,j}\alpha_j^\ast)
		=\beta_{i},
		\\
		\langle \hat b_{i}\hat b_{i'} \rangle
		=&\beta_{i}\beta_{i'}+z_{i,i'},
		\\
		\langle \hat b_{i}^\dag\hat b_{i'} \rangle
		=&\beta_{i}^\ast\beta_{i'}+y_{i,i'},
		\\
		\langle \hat b_{i}^\dag\hat b_{i'}^\dag\hat b_{i}\hat b_{i'} \rangle
		=&|z_{i,i'}|^2
		+y_{i,i}y_{i',i'}
		+y_{i,i'}y_{i',i}
		+z_{i,i'}^\ast\beta_{i}\beta_{i'}
		\\
		&+z_{i,i'}\beta_{i}^\ast\beta_{i'}^\ast
		+y_{i,i}|\beta_{i'}|^2
		+y_{i',i'}|\beta_{i}|^2
		\\
		&+y_{i,i'}\beta_{i'}^\ast\beta_{i}
		+y_{i',i}\beta_{i}^\ast\beta_{i'}
		+|\beta_{i}|^2|\beta_{i'}|^2,
	\end{aligned}
	\end{equation}
	using the output coherent amplitude $\beta_i$ for brevity although it is completely expressed in terms of the input state (see the first line of the equation).
	We then also find the central field moments as $\langle (\Delta\hat b_{i})(\Delta\hat b_{i'}) \rangle=z_{i,i'}$ and $\langle (\Delta\hat b_{i})^\dag(\Delta\hat b_{i'}) \rangle=y_{i,i'}$.
	Furthermore, normally ordered photon-number moments are
	\begin{equation}
	\begin{aligned}
		\langle \hat n_{i} \rangle=&|\beta_{i}|^2+y_{i,i}
		\\
		\langle {:} (\Delta\hat n_{i})(\Delta\hat n_{i'}) {:} \rangle
		=&|z_{i,i'}|^2
		+|y_{i,i'}|^2
		+z_{i,i'}^\ast\beta_{i}\beta_{i'}
		\\
		&+z_{i,i'}\beta_{i}^\ast\beta_{i'}^\ast
		+y_{i,i'}\beta_{i}^\ast\beta_{i'}
		+y_{i,i'}^\ast\beta_{i'}^\ast\beta_{i}.
	\end{aligned}
	\end{equation}
	For quadrature and momentum moments and their covariances, we analogously find the relations $\langle \hat q_i\rangle=\sqrt{2}\mathrm{Re}(\beta_i)$ and $\langle \hat p_i\rangle=\sqrt{2}\mathrm{Im}(\beta_i)$, as well as
	\begin{equation}
	\begin{aligned}
		\frac{1}{2}\langle \{\Delta \hat q_{i},\Delta \hat q_{i'}\} \rangle
		=&\mathrm{Re}(y_{i,i'})+\mathrm{Re}(z_{i,i'})+\frac{1}{2}\delta_{i=i'},
		\\
		\frac{1}{2}\langle \{\Delta \hat p_{i},\Delta \hat p_{i'}\} \rangle
		=&\mathrm{Re}(y_{i,i'})-\mathrm{Re}(z_{i,i'})+\frac{1}{2}\delta_{i=i'},
		\\
		\frac{1}{2}\langle \{\Delta \hat q_{i},\Delta \hat p_{i'}\} \rangle
		=&\mathrm{Im}(y_{i,i'})+\mathrm{Im}(z_{i,i'}).
	\end{aligned}
	\end{equation}

\section{Propagation}
\label{App:Propagators}

	Our scenario of a one-dimensional quantum walk with periodic boundaries consists in both cases of the step operation $\Sigma_{\pm}$, Eq. \eqref{eq:DefineSigmaPlusMinus}.
	In particular, a step $x\mapsto x+1\,\mathrm{mod}\,d$ is described by the operator $\Sigma$ in Eq. \eqref{eq:DefineSigma}.
	This map obeys $\Sigma=\Sigma^\ast$ and $\Sigma^{-1}=\Sigma^\mathrm{T}$.
	The step $x\mapsto x-1\,\mathrm{mod}\,d$ is given through $\Sigma^\mathrm{T}$.
	The two distinct coin operations can be expressed via the matrices $\Gamma_\mathrm{BS}$ [Eq. \eqref{eq:DefGammaBS}] and $\Gamma_\mathrm{PDC}$ [Eq. \eqref{eq:DefGammaPDC}].
	Both include submatrices that obey $1
		=\det\left[\begin{smallmatrix}\tau&\rho\\-\rho^\ast&\tau^\ast\end{smallmatrix}\right]
		=\det\left[\begin{smallmatrix}\tau^\ast&\rho^\ast\\-\rho&\tau\end{smallmatrix}\right]
		=\det\left[\begin{smallmatrix}\mu&\nu\\\nu^\ast&\mu^\ast\end{smallmatrix}\right]
	$, showing that the two eigenvalues are reciprocals to each other.
	Furthermore, a combination of coin and step operation yields a single time step of the walk protocol,
	\begin{equation}
		T_{1,\mathrm{BS}}=\begin{bmatrix}
			\tau\Sigma & \rho\Sigma & 0 & 0
			\\
			-\rho^\ast\Sigma^\mathrm{T} & \tau^\ast\Sigma^\mathrm{T} & 0 & 0
			\\
			0 & 0 & \tau^\ast\Sigma & \rho^\ast\Sigma
			\\
			0 & 0 & -\rho\Sigma^\mathrm{T} & \tau\Sigma^\mathrm{T}
		\end{bmatrix}
	\end{equation}
	and
	\begin{equation}
		T_{1,\mathrm{PDC}}=\begin{bmatrix}
			\mu\Sigma & 0 & 0 & \nu\Sigma
			\\
			0 & \mu\Sigma^\mathrm{T} & \nu\Sigma^\mathrm{T} & 0
			\\
			0 & \nu^\ast\Sigma & \mu^\ast\Sigma & 0
			\\
			\nu^\ast\Sigma^\mathrm{T} & 0 & 0 & \mu^\ast\Sigma^\mathrm{T}
		\end{bmatrix}.
	\end{equation}
	Applying a spectral decomposition, the target is now to express the evolution matrix $T$ after $t$ time steps, i.e., $T=T_1^t$, for both families of coins.
	Similarly, for $G=\ln T_1$, we have $T=\exp(tG)$ for the continuous case, with $T=\mathrm{id}$ for $t=0$.

	It is known that the matrix $\Sigma$ can be diagonalized using a discrete Fourier transform, $F=[\omega^{kx}/\sqrt{d}]_{k,x=0,\ldots,d-1}=F^\mathrm{T}$, with the constant $\omega=\exp(2\pi i/d)$ and the inverse $F^{-1}=F^{\ast\mathrm{T}}=F^{\ast}$.
	That is, we can write
	\begin{equation}
		\Sigma=F^\ast\Omega F,
		\quad\text{where}\quad
		\Omega=\mathrm{diag}(\omega^0,\ldots,\omega^{d-1}).
	\end{equation}
	Note that $\Sigma=\Sigma^\ast$ holds true;
	hence, we can equivalently write $\Sigma=F\Omega^\ast F^\ast$.
	For the transposed step matrix, we have $\Sigma^\mathrm{T}=F\Omega F^\ast=F^\ast\Omega^\ast F$.
	Those identities enable us to rewrite our single-step transformations as
	\begin{equation}
	\begin{aligned}
		T_{1,\mathrm{BS}}
		=&\mathrm{diag}(F^\ast,F^\ast,F,F)
		\\&\times
		\begin{bmatrix}
			\tau \Omega & \rho \Omega & 0 & 0
			\\
			-\rho^\ast \Omega^\ast & \tau^\ast \Omega^\ast & 0 & 0
			\\
			0 & 0 & \tau^\ast\Omega^\ast & \rho^\ast \Omega^\ast
			\\
			0 & 0 & -\rho\Omega & \tau\Omega
		\end{bmatrix}
		\\&\times
		\mathrm{diag}(F,F,F^\ast,F^\ast)
	\end{aligned}
	\end{equation}
	and
	\begin{equation}
	\begin{aligned}
		T_{1,\mathrm{PDC}}
		=&\mathrm{diag}(F^\ast,F,F,F^\ast)
		\\&\times
		\begin{bmatrix}
			\mu \Omega & 0 & 0 & \nu\Omega
			\\
			0 & \mu \Omega & \nu \Omega & 0
			\\
			0 & \nu^\ast\Omega^\ast & \mu^\ast\Omega^\ast & 0
			\\
			\nu^\ast\Omega^\ast & 0 & 0 & \mu^\ast\Omega^\ast
		\end{bmatrix}
		\\&\times
		\mathrm{diag}(F,F^\ast,F^\ast,F).
	\end{aligned}
	\end{equation}
	It is noteworthy that the block diagonals that include the Fourier matrices have different orders of those matrices and their conjugates ones for both coin types.
	Because the central matrix between the Fourier matrices includes only diagonal blocks, we can now focus on the submatrices of the form
	\begin{equation}
		\tilde T_{1,\mathrm{BS}}=\begin{bmatrix}
			\tilde\tau & \tilde\rho & 0 & 0
			\\
			-\tilde\rho^\ast & \tilde\tau^\ast & 0 & 0
			\\
			0 & 0 & \tilde\tau^\ast & \tilde\rho^\ast
			\\
			0 & 0 & -\tilde\rho & \tilde\tau
		\end{bmatrix}
	\end{equation}
	and
	\begin{equation}
		\tilde T_{1,\mathrm{PDC}}=\begin{bmatrix}
			\tilde \mu & 0 & 0 & \tilde \nu
			\\
			0 & \tilde \mu & \tilde \nu & 0
			\\
			0 & \tilde \nu^\ast & \tilde \mu^\ast & 0
			\\
			\tilde \nu^\ast & 0 & 0 & \tilde \mu^\ast
		\end{bmatrix},
	\end{equation}
	where $\tilde \mu=\omega^k\mu$, $\tilde \nu=\omega^k\nu$, $\tilde \tau=\omega^k\tau$, and $\tilde \rho=\omega^k\rho$ for any wave number $k\in\{0,\ldots,d-1\}$, while still satisfying $|\tilde\tau|^2+|\tilde \rho|^2=1$ and $|\tilde\mu|^2-|\tilde\nu|^2=1$.
	The simplified $4\times4$ single-time-step matrices $\tilde T_{1,\mathrm{BS}}$ and $\tilde T_{1,\mathrm{PDC}}$ further decompose into $2\times 2$ submatrices that are given by the invariant subspaces, easily identifiable via zero entries.

	For arbitrary $2\times 2$ matrices $\tilde M$ and arbitrary analytic functions $f$, this function acts on the matrix like
	\begin{equation}
	\begin{aligned}
		f(\tilde M)
		=&\frac{f(\tilde W+\varepsilon)+f(\tilde W-\varepsilon)}{2}\mathrm{id}
		\\
		&+\frac{f(\tilde W+\varepsilon)-f(\tilde W-\varepsilon)}{2\varepsilon}\left(\tilde M-\tilde W\mathrm{id}\right),
	\end{aligned}
	\end{equation}
	with
	\begin{equation}
	\label{eq:2x2parameters}
		\varepsilon=\sqrt{\tilde W^2-\det(\tilde M)}
		\quad\text{and}\quad
		\tilde W=\frac{\mathrm{tr}(\tilde M)}{2},
	\end{equation}
	which is a consequence of the Taylor expansion of $f$ and the fact that $[\tilde M-\tilde W\mathrm{id}]^2=\varepsilon^2\mathrm{id}$ (i.e., a multiple of the identity) holds true for all $2\times 2$ matrices.
	In addition, we have a removable singularity at $\epsilon=0$,
	\begin{equation}
		\lim_{\epsilon\to0}f(\tilde M)
		=f(\tilde W)\mathrm{id}
		+f'(\tilde W)(\tilde M- \tilde W\mathrm{id}),
	\end{equation}
	where $f'$ is the derivative of $f$;
	this applies if $\tilde M$ has degenerate eigenvalues.
	The broader treatment described here is helpful for our case where $f$ describes the $t$th power of $\tilde M$.

	Please further note that the involved arguments $\tilde W\pm\varepsilon$ are the eigenvalues of $\tilde M$.
	And, if $\det\tilde M=1$ applies, both eigenvalues are reciprocals to each other because of $(\tilde W\pm\varepsilon)^{-1}=(\tilde W\mp\varepsilon)/\det(\tilde M)$.
	Since this is true in our scenarios, $(\tilde W-\varepsilon)^t=(\tilde W+\varepsilon)^{-t}$ too holds true.
	We can additionally use the simple identity $c^t=\exp(t\ln c)$ for complex numbers $c$.

	We now apply all above considerations to our invariant subspaces of the $4\times4$ matrices.
	For our beam splitter coin, we obtain
	\begin{equation}
	\begin{aligned}
		(\tilde T_{1,\mathrm{BS}})^t
		=&\cos\left[t\arg\left(\mathrm{Re}\,\tilde\tau+i\sqrt{1-(\mathrm{Re}\,\tilde\tau)^2}\right)\right]
		\mathrm{id}
		\\
		&+\frac{\sin\left[t\arg\left(\mathrm{Re}\,\tilde\tau+i\sqrt{1-(\mathrm{Re}\,\tilde\tau)^2}\right)\right]}{\sqrt{1-(\mathrm{Re}\,\tilde\tau)^2}}
		\\
		&\phantom{ + }\times
		\begin{bmatrix}
			i\mathrm{Im}\,\tilde\tau & \tilde\rho & 0 & 0
			\\
			-\tilde\rho^\ast & -i\mathrm{Im}\,\tilde\tau & 0 & 0
			\\
			0 & 0 & -i\mathrm{Im}\,\tilde\tau & \tilde\rho^\ast
			\\
			0 & 0 & -\tilde\rho & i\mathrm{Im}\,\tilde\tau
		\end{bmatrix}
	\end{aligned}
	\end{equation}
	after some straightforward algebra that utilizes the discussion above and the identity $\mathrm{Re}\,\tilde\tau-i\sqrt{1-(\mathrm{Re}\,\tilde\tau)^2}=[\mathrm{Re}\,\tilde\tau+i\sqrt{1-(\mathrm{Re}\,\tilde\tau)^2}]^{-1}$.
	Analogously, the two-mode squeezing coin can be analyzed, yielding
	\begin{equation}
	\begin{aligned}
		(\tilde T_{1,\mathrm{PDC}})^t
		=&\cosh\left[t\ln\left(\mathrm{Re}\,\tilde\tau+\sqrt{(\mathrm{Re}\,\tilde\tau)^2-1}\right)\right]
		\mathrm{id}
		\\
		&+\frac{\sinh\left[t\ln\left(\mathrm{Re}\,\tilde\tau+\sqrt{(\mathrm{Re}\,\tilde\tau)^2-1}\right)\right]}{\sqrt{(\mathrm{Re}\,\tilde\tau)^2-1}}
		\\
		&\phantom{ + }\times
		\begin{bmatrix}
			i\mathrm{Im}\,\tilde \mu & 0 & 0 & \tilde \nu
			\\
			0 & i\mathrm{Im}\,\tilde \mu & \tilde \nu & 0
			\\
			0 & \tilde \nu^\ast & -i\mathrm{Im}\,\tilde \mu & 0
			\\
			\tilde \nu^\ast & 0 & 0 & -i\mathrm{Im}\,\tilde \mu
		\end{bmatrix},
	\end{aligned}
	\end{equation}
	with $\mathrm{Re}\,\tilde\mu-\sqrt{(\mathrm{Re}\,\tilde\mu)^2-1}=[\mathrm{Re}\,\tilde\mu-\sqrt{(\mathrm{Re}\,\tilde\tau)^2-1}]^{-1}$.
	Please note that the square root can become a complex number in some instances.
	See the discussion in Sec. \ref{Subsec:SpectralSolve} for the physical interpretation.

\begin{widetext}
	In conclusion, the $t$-time-step evolution can be analytically described as follows.
	Firstly, we define diagonal matrices
	\begin{equation}
		\Theta=\arg\left\{\mathrm{Re}(\tau\Omega)+i\sqrt{\mathrm{id}-[\mathrm{Re}(\tau\Omega)]^2}\right\}
		\quad\text{and}\quad
		\Xi=\ln\left\{\mathrm{Re}(\mu\Omega)+\sqrt{[\mathrm{Re}(\mu\Omega)]^2-\mathrm{id}}\right\},
	\end{equation}
	where all involved functions are considered to act entrywise on matrices.
	Then, we can write for $T=(T_{1})^t$ and the two coin types
	\begin{equation}
	\begin{aligned}
		T_\mathrm{BS}
		=&
		\mathrm{diag}(F^\ast,F^\ast,F,F)
		\\
		&\times
		\left\{
			\cos(t\Theta)
			\begin{bmatrix}
				\mathrm{id} & 0 & 0 & 0
				\\
				0 & \mathrm{id} & 0 & 0
				\\
				0 & 0 & \mathrm{id} & 0
				\\
				0 & 0 & 0 & \mathrm{id}
			\end{bmatrix}
			+\frac{\sin(t\Theta)}{\sqrt{\mathrm{id}-[\mathrm{Re}(\tau\Omega)]^2}}
			\begin{bmatrix}
				i\mathrm{Im}(\tau\Omega) & \rho\Omega & 0 & 0
				\\
				-\rho^\ast\Omega^\ast & -i\mathrm{Im}(\tau\Omega) & 0 & 0
				\\
				0 & 0 & -i\mathrm{Im}(\tau\Omega) & \rho^\ast\Omega^\ast
				\\
				0 & 0 & -\rho\Omega & i\mathrm{Im}(\tau\Omega)
			\end{bmatrix}
		\right\}
		\\
		&\times
		\mathrm{diag}(F,F,F^\ast,F^\ast)
	\end{aligned}
	\end{equation}
	and
	\begin{equation}
	\begin{aligned}
		T_\mathrm{PDC}
		=&
		\mathrm{diag}(F^\ast,F,F,F^\ast)
		\\
		&\times
		\left\{
			\cosh(t\Xi)
			\begin{bmatrix}
				\mathrm{id} & 0 & 0 & 0
				\\
				0 & \mathrm{id} & 0 & 0
				\\
				0 & 0 & \mathrm{id} & 0
				\\
				0 & 0 & 0 & \mathrm{id}
			\end{bmatrix}
			+\frac{\sinh(t\Xi)}{\sqrt{[\mathrm{Re}(\mu\Omega)]^2-\mathrm{id}}}
			\begin{bmatrix}
				i\mathrm{Im}(\mu\Omega) & 0 & 0 & \nu\Omega
				\\
				0 & i\mathrm{Im}(\mu\Omega) & \nu\Omega & 0
				\\
				0 & \nu^\ast\Omega^\ast & -i\mathrm{Im}(\mu\Omega) & 0
				\\
				\nu^\ast\Omega^\ast & 0 & 0 & -i\mathrm{Im}(\mu\Omega)
			\end{bmatrix}
		\right\}
		\\
		&\times
		\mathrm{diag}(F,F^\ast,F^\ast,F),
	\end{aligned}
	\end{equation}
	where the (hyperbolic) cosine and sine matrices have to be understood as matrix-valued factors for each matrix-valued entry of the following $4\times 4$ block matrix.
	(Since all involved matrices are diagonal, all matrix products within the curled brackets commute.)
	These analytic results are used for the simulation of the quantum walks studied in the main text, including their time-dependent properties, encoded in $T$, such as entanglement, photon-number correlations, amplification, etc.

\section{Entanglement test}
\label{App:Entanglement}

	For self-consistency, we recall the known results for our entanglement test in terms of quadratures and momenta \cite{SV13}.
	Here, however, we shall formulate the results in terms of bosonic ladder operators.
	Our test operator from Eq. \eqref{eq:TestOp} expands as
	\begin{equation}
	\begin{aligned}
		\hat L
		=&\sum_{j,j'}\left(
			\tilde u_{j,j'}\hat b_{j}^\dag\hat b_{j'}
			+\tilde v_{j,j'}\hat b_{j}\hat b_{j'}^\dag
			-\tilde w_{j,j'}\hat b_{j}^\dag\hat b_{j'}^\dag
			-\tilde w_{j',j}^\ast\hat b_{j}\hat b_{j'}
		\right),
	\end{aligned}
	\end{equation}
	where $\tilde u_{j,j'}=\sum_{i}U_{j,i}U^\ast_{j',i}$, $\tilde v_{j,j'}=\sum_{i}V^\ast_{j,i}V_{j',i}$, and $\tilde w_{j,j'}=\sum_iU_{j,i}V_{j',i}$ are introduced for convenience.
	The goal now is to determine the minimal expectation value for states that are separated in the output modes, $\bigotimes_j|\psi_j\rangle$.

	For that purpose, we compute the so-called partially reduced operator for the $j_0$th subsystem by tracing over the remaining states \cite{SV13}, which is denoted by and expanded as
	\begin{equation}
	\begin{aligned}
		\hat L_{\ldots,\psi_{j_0-1},\psi_{j_0+1},\ldots}
		=&\left(
			\tilde u_{j_0,j_0}\hat b_{j_0}^\dag\hat b_{j_0}
			+\tilde v_{j_0,j_0}\hat b_{j_0}\hat b_{j_0}^\dag
			-\tilde w_{j_0,j_0}\hat b_{j_0}^\dag\hat b_{j_0}^\dag
			-\tilde w_{j_0,j_0}^\ast\hat b_{j_0}\hat b_{j_0}
		\right)
		\\
		&+\sum_{j\neq j_0}\left(
			\tilde u_{j,j}\langle\hat b_{j}^\dag\hat b_{j}\rangle_{\not j_0}
			+\tilde v_{j,j}\langle\hat b_{j}\hat b_{j}^\dag\rangle_{\not j_0}
			-\tilde w_{j,j}\langle\hat b_{j}^\dag\hat b_{j}^\dag\rangle_{\not j_0}
			-\tilde w_{j,j}^\ast\langle\hat b_{j}\hat b_{j}\rangle_{\not j_0}
		\right)
		,
	\end{aligned}
	\end{equation}
	where we used the abbreviation $\langle \cdots\rangle_{\not j_0}$ to indicate expectation values that do not include the $j_0$th component.
	Moreover, we also utilized the fact that the minimal eigenvalue is attained for $\langle \hat b_j\rangle=0$ for all $j$s \cite{GSVCRTF15}.
	While the second part of the partially reduced operator, which includes the sum, is a constant with respect to $j_0$, the first contribution resembles a single-mode squeezing Hamiltonian.
	The known minimal expectation value of its (squeezed) ground state $|\psi_{j_0}\rangle$ is
	\begin{equation}
		\langle\psi_{j_0}|
			\tilde u_{j_0,j_0}\hat b_{j_0}^\dag\hat b_{j_0}
			+\tilde v_{j_0,j_0}\hat b_{j_0}\hat b_{j_0}^\dag
			-\tilde w_{j_0,j_0}\hat b_{j_0}^\dag\hat b_{j_0}^\dag
			-\tilde w_{j_0,j_0}^\ast\hat b_{j_0}\hat b_{j_0}
		|\psi_{j_0}\rangle
		=\frac{
			\tilde v_{j_0,j_0}-\tilde u_{j_0,j_0}
			+\sqrt{(\tilde v_{j_0,j_0}+\tilde u_{j_0,j_0})^2-4|\tilde w_{j_0,j_0}|^2}
		}{2},
	\end{equation}
	which holds true for all $j_0$s.
	Therefore, the minimal expectation value for the full tensor-product states reads
	\begin{equation}
	\begin{aligned}
		g_{\min}=&\langle\psi_{j_0}| \hat L_{\ldots,\psi_{j_0-1},\psi_{j_0+1},\ldots} |\psi_{j_0}\rangle
		=\left(\bigotimes_j\langle\psi_j|\right)
		\hat L
		\left(\bigotimes_j|\psi_j\rangle\right)
		\\
		=&\frac{1}{2}\sum_j\left(
			\tilde v_{j,j}-\tilde u_{j,j}
			+\sqrt{(\tilde v_{j,j}+\tilde u_{j,j})^2-4|\tilde w_{j,j}|^2}
		\right)
		=\frac{1}{2}\sum_j\left(
			\sqrt{(1+2\tilde v_{j,j})^2-4|\tilde w_{j,j}|^2}
			-1
		\right),
	\end{aligned}
	\end{equation}
	where the relation $\tilde u_{j,j}=1+\tilde v_{j,j}$---being a consequence of Eq. \eqref{eq:UVconditions}---is applied to express the final formula.
\end{widetext}


\end{document}